\documentclass[aps,pre,
    tightenlines,
   reprint,
   superscriptaddress,
   eqsecnum,
   floatfix,
   showpacs,
   preprintnumbers,
   nofootinbib
   ]{revtex4-1}
\usepackage{color}
\usepackage{framed}
\usepackage{subfigure}
\usepackage{time}
\usepackage{graphicx}
\DeclareSymbolFont{AMSb}{U}{msb}{m}{n}
\DeclareSymbolFontAlphabet{\mathbb}{AMSb}
\usepackage{physics}                           
\newcommand{\ef}[1]{\eqref{#1}}                

\newcommand{\mc}[1]{\mathcal{#1}}              
\newcommand{\PT}{\mathcal{PT}}                 

\newcommand{\bq}{\begin{equation}}
\newcommand{\ee}{\end{equation}}
\newcommand{\bea}{\begin{eqnarray}}
\newcommand{\eea}{\end{eqnarray}}

\newcommand{\ba}{\begin{eqnarray}}
\newcommand{\eq}{\end{equation}}
\newcommand{\ea}{\end{eqnarray}}

\newcommand{\PB}[2]{\ensuremath{ \lbrace \,#1,#2\, \rbrace }}  
\newcommand{\tint}{\!\int\!}                   
\newcommand{\tintx}{\!\int\!\mathrm{d}x\,}     
\newcommand{\tinty}{\!\int\!\mathrm{d}y\,}     

\newcommand{\rmi}{{\rm i}}                     
\newcommand{\rme}{{\rm e}}                     

\newcommand{\tpsi}{\tilde{\psi}}               

\newcommand{\R}{\mathbb{R}}

\newcommand{\vs}{vs.}                      
\newcommand{\ie}{\textit{i.e.}}            
\newcommand{\Schrodinger}{Schr{\"o}dinger}
\newcommand{\tphi} {{\tilde \phi}}
\newcommand{\Comm}[2]%
   {\ensuremath{[ \, #1, #2 \, ]}}
\newcommand{\MEangle}[3]%
   {\ensuremath{\langle \, #1 \, | \, #2 \, | \, #3 \, \rangle}}
\usepackage{hyperref}
\hypersetup{
    unicode=false,          
    pdftoolbar=true,        
    pdfmenubar=true,        
    pdffitwindow=false,     
    pdfstartview={FitH},    
    pdftitle={Paper},       
    pdfauthor={John Dawson},     
    pdfsubject={Phys Rev article},   
    pdfcreator={John Dawson},   
    pdfproducer={John Dawson}, 
    pdfkeywords={keyword1} {key2} {key3}, 
    pdfnewwindow=true,      
    colorlinks=true,        
    linkcolor=red,          
    citecolor=blue,         
    filecolor=green,        
    urlcolor=cyan           
}
\begin{document}
%
%
\preprint{LA-UR-17-25332}
\title{Response of exact solutions of the nonlinear \Schrodinger\ equation to small perturbations in a class of complex external potentials having supersymmetry and parity-time symmetry } 
\author{Fred Cooper} 
\email{cooper@santafe.edu}
\affiliation{The Santa Fe Institute, 1399 Hyde Park Road, Santa Fe, NM 87501, USA}
\affiliation{Theoretical Division,
   Los Alamos National Laboratory,
   Los Alamos, NM 87545, USA}
\author{John F. Dawson}
\email{john.dawson@unh.edu}
\affiliation{Department of Physics,
   University of New Hampshire,
   Durham, NH 03824, USA}   
\author{Franz G. Mertens}
\email{Franz.Mertens@uni-bayreuth.de}
\affiliation{Physikalisches Institut, Universit{\"a}t Bayreuth, D-95440 Bayreuth, Germany} 
\author{Edward Ar\'evalo}
\email{earevalo@fis.puc.cl} 
\affiliation{Pontifical Catholic University of Chile,
{ Instituto} de F\'{\i}sica, Santiago, Regi\'on Metropolitana, Chile}
\author{Niurka R. Quintero}
\email{niurka@us.es}
\affiliation{Instituto Carlos I de Fisica Te{\'o}rica y Computacional,
Universidad de Granada, E-18015, Granada, Spain}
\affiliation{Departamento de Fisica Aplicada I, 
E.P.S. Universidad de Sevilla, 41011 Sevilla, Spain}
\author{Bogdan Mihaila}
\email{bmihaila05@gmail.com}
\affiliation{Physics Division, National Science Foundation, Arlington, VA 22230, USA}
\author{Avinash Khare}
\email{khare@physics.unipune.ac.in}
\affiliation{Physics Department, Savitribai Phule Pune University, Pune 411007, India}
\author{Avadh Saxena} 
\email{avadh@lanl.gov}
\affiliation{Theoretical Division,
   Los Alamos National Laboratory,
   Los Alamos, NM 87545, USA}
%
%
\date{\today, \now \ EDT}
%
%
%
\begin{abstract}
We  discuss the effect of small perturbation on nodeless solutions of the nonlinear \Schrodinger\  equation in 1+1 dimensions in an external complex potential derivable from a parity-time symmetric superpotential that was considered earlier [Phys.~Rev.~E 92, 042901 (2015)].  In particular we consider the nonlinear partial differential equation 
$\{ \, \rmi \, \partial_t + \partial_x^2 + g |\psi(x,t)|^2 - V^{+}(x) \, \} \, \psi(x,t) = 0$, 
where $V^{+}(x) = \qty( -b^2 - m^2 + 1/4 ) \, \sech^2(x) - 2 i \, m \, b \, \sech(x) \, \tanh(x)$ represents the complex potential.
Here we study the perturbations as a function of $b$ and $m$ using a variational approximation based on a dissipation functional formalism.  We compare the result of this variational approach  with direct numerical simulation of the equations. We find that the variational approximation works quite well at small and moderate values of the parameter $b m$ which controls the strength of the imaginary part of the potential. We also show that the dissipation functional formalism is equivalent to the generalized traveling wave method for this type of dissipation.
\end{abstract}
\maketitle

%
%
\section{\label{s:Intro}Introduction}

The topic of balanced loss and gain or parity-time ($\PT$) symmetry and its relevance for physical 
applications on the one hand, as well as its mathematical structure on the other, have drawn considerable 
attention from both the physics and the mathematics community.  The original proposal of Bender and his 
collaborators \cite{r:Bender:2007nr,0305-4470-39-32-E01,1751-8121-41-24-240301,1751-8121-45-44-440301} 
towards the study of such systems was made as an alternative to the postulate of Hermiticity in 
quantum mechanics.  Keeping in perspective the formal similarity of the \Schrodinger\ equation with 
Maxwell's equations in the paraxial approximation, it was realized that such $\PT$ invariant systems can in fact be experimentally realized in optics \cite{Makris2011,0305-4470-38-9-L03,PhysRevLett.100.103904,PhysRevLett.101.080402,PhysRevLett.103.123601,PhysRevB.80.235102,PhysRevA.81.022102,r:Ruter:2010mz,PhysRevLett.103.093902,r:Regensburger:2012gf}. 
Subsequently, these efforts motivated experiments in several other areas including $\mc{PT}$ invariant electronic circuits \cite{PhysRevA.84.040101,1751-8121-45-44-444029}, mechanical circuits \cite{r:Bender:2013ly}, and whispering-gallery microcavities \cite{r:Peng:2014ul}.

Concurrently, the notion of supersymmetry (SUSY) originally espoused in high-energy physics has also been realized in optics \cite{PhysRevLett.110.233902,r:Heinrich:2014qf}.  The key idea is that from a given potential one can obtain a SUSY partner potential with both potentials possessing the same spectrum, except possibly for one eigenvalue \cite{0038-5670-28-8-R01,r:CooperKhareSukhatmePR}.  Therefore, an interplay of SUSY with $\mc{PT}$ symmetry is expected to be quite rich and is indeed very useful in achieving transparent as well as one-way reflectionless complex optical potentials \cite{0305-4470-33-1-101,doi:10.1142/S0217751X01004153,Bagchi2000285,Ahmed2001343,PhysRevA.89.032116}.  

A previous paper \cite{PhysRevE.92.042901} explored the interplay between $\mc{PT}$ symmetry, SUSY and nonlinearity.  That paper derived exact solutions of the general nonlinear \Schrodinger\ (NLS) equation in 1+1 dimensions when in a $\PT$-symmetric complex potential \cite{0038-5670-28-8-R01,r:CooperKhareSukhatmeBOOK}.
In particular, they considered the nonlinear partial differential equation
\begin{equation}\label{eqn1}
   \qty{
   i \,
   \partial_t
   +  
   \partial_x^2
   - 
   V^\pm(x)
   +
   g | \psi(x,t) |^{2\kappa}
   } \, \psi(x,t) 
   = 
   0 \>,
\end{equation}
for arbitrary nonlinearity parameter $\kappa$, with
\begin{equation}\label{eqn1x}
   V^{\pm}(x)
   =
   W_1^2(x)\mp W_1'(x) - (m-1/2)^2 \>,
\end{equation}
and the partner potentials arise from the superpotential 
\begin{equation}\label{eqn1y}
   W_1(x)
   =
   \qty( m - 1/2 ) \,  \tanh{x} 
   - 
   i b \, \sech{x} \>,
\end{equation}
giving rise to
\begin{subequations}\label{VpVm}
\begin{align}
   V^{+}(x)
   &= 
   \qty( -b^2 - m^2 + 1/4 ) \, \sech^2(x) 
   \label{eqn7a} \\
   & \quad
   - 
   2 i \, m \, b \, \sech(x) \, \tanh(x),
   \notag \\
   V^{-}(x)
   &=
   \qty( - b^2 - (m-1)^2 + 1/4 ) \, \sech^2(x)
   \label{eqn8a} \\
   & \quad
   -
   2 i \, \qty( m - 1 ) \, b \, \sech(x) \, \tanh(x) \>.
   \notag
\end{align}
\end{subequations}
For $m=1$, the \emph{complex} potential $V^{+}(x)$ has the same spectrum, apart from the ground state, as the \emph{real} potential $V^{-}(x)$ and this fact was used in the numerical study of the stability of the bound state solutions of the NLS equation in the presence of $V^{+}(x)$ (see Ref.~\cite{PhysRevE.92.042901}).
In a recent complementary study \cite{Cooper:2017aa} of this system of nonlinear \Schrodinger\ equations in $\mc{PT}$ symmetric SUSY external potentials, the stability properties of the bound state solutions of NLS equation in the presence of the external real SUSY partner potential $V^{-}(x)$ were investigated.  The stability regime of these solutions, which depended on the parameters $(b, \kappa)$, was compared to the stability regime of the related \emph{solitary} wave solutions to the NLS equation in the absence of the external potential.  Because the NLS equation in the presence of $V^{-}(x)$ is a Hamiltonian dynamical system, in Ref.~\cite{Cooper:2017aa} they were able to use several variational methods to study the stability of the solutions when they undergo certain small deformations, and showed that these variational methods agreed with a linear stability analysis based on the Vakhitov-Kolokolov (V-K) stability criterion \cite{r:Comech:2012uk,Vakhitov:1973aa} as well as numerical simulations that have recently been performed.

In Ref.~\cite{PhysRevE.92.042901} we determined the exact solutions of the equation for $m=1$ for $V^{+}(x)$, which was complex. We studied numerically the stability properties of these solutions using linear stability analysis.  We found some unusual results for the stability which depended on the value of $b$.  What was found for $m=1$ (and $\kappa=1$) was that the eigenvalues of the linear stability matrix became complex for $0.56 < b < 1.37$. 

At that time we had not yet formulated a variational approach for deriving the NLS equation in the presence of complex potentials.  Recently we have developed such an approach and have applied it to the response of the solutions of the NLS equation to  weak external complex periodic potentials.  Using four variational parameters we were able to successfully predict the time evolution of these solitary waves  when compared to direct numerical simulation of the NLS equation in the presence of these complex potentials \cite{PhysRevE.94.032213}. Given this new tool we would like to return to the original problem of the stability of the exact solutions found in Ref.~\cite{PhysRevE.92.042901} and see how well this variational approach agrees with numerical simulations as a function of the strength of the dissipative part of the potential which is proportional to $bm$. 
In this paper we focus on the external potential $V^{+}(x)$ which is symmetric in $b \leftrightarrow m$.

Here we will compare the numerical simulations with the results of our collective coordinate (CC) approximation.  We will also look at the linear stability analysis that arises from studying the linearization of the CC ordinary differential equations (ODEs).  For the case of a real external potential, studying the eigenvalues of this reduced stability analysis predicted the correct stability regime \cite{PhysRevE.85.046607}. 

This paper is structured as follows.  In Sec.~\ref{s:SSModel} we review the non-hermitian SUSY model that we studied in Ref.~\cite{PhysRevE.92.042901} and add the self-interactions of the NLS equation to the linear model. In Sec.~\ref{s:general} we give some of the exact low order moment equations for this problem.  In Sec.~\ref{s:CCs} we introduce our collective coordinate approach, whereas in Sec.~\ref{s:FourPar} we use a four parameter trial wave function that we considered in an earlier study of soliton behavior in complex periodic external potentials, and derive equations for the four CC's.  In Sec.~\ref{s:6cc} we expand the number of CC's to six and derive equations for the six CC's.  In Sec.~\ref{s:LinearRes} we study the linear response theory of the six CC approximation.  In Sec.~\ref{s:Numerical} we present our numerical strategy for solving the NLS equation starting from a perturbed exact solution.  In Sec.~\ref{s:results} we compare the four and  six CC approximations with direct numerical simulations.  In Sec.~\ref{s:conclusions} we present our main conclusions.  Finally in Appendix A we provide the definitions of various integrals and in Appendix B we show that for this problem our variational approach is equivalent to the generalized traveling wave method \cite{PhysRevE.82.016606}.
%
%
\section{\label{s:SSModel}NLS equation in the presence of a  Non-Hermitian Supersymmetric external potential}

We were interested in studying the NLS equation in the presence of a complex external potential and were intrigued by the fact that as a result of $\mathcal{PT}$ symmetry, there existed complex potentials whose SUSY partners were real and had explicitly known spectra of bound states. This led us to study the external potential defined by the $\mathcal{PT}$ symmetric SUSY superpotential $W_1(x)$ given by Eq. \eqref{eqn1y}.
This superpotential gives rise to supersymmetric partner potentials given by Eqs.~\ef{VpVm}.  For the case $m=1$, $V^{-}(x)$ is the well known P{\"o}schl-Teller potential \cite{r:Poschl:1933ek,r:Landau:1989jt}.  The relevant bound state eigenvalues assume an extremely simple form as  
\begin{equation}\label{eqn9}
   E_n^{(-)}
   =
   -\frac{1}{4} \, \qty[ \, 2 b - 2 n - 1 \, ]^2 \>.
\end{equation}
Such bound state eigenvalues only exist when $n < b - 1/2$.  We notice that for  the ground state (n=0) to exist requires $b > 1/2$.  The existence of a first excited state (n=1) requires $b > 3/2$.
Here we consider the  general   $V^{+}(x)$  arising from the superpotential $W_1(x)$ depending on $m,b$ as an external potential modifying the nonlinear \Schrodinger\ equation.  Rewriting the external potential given in Eq.~\ef{eqn7a} as,
\begin{equation}\label{e:Vp}
   V^{+}(x)
   = 
   V_1(x) + \rmi \, V_2(x) \>,
\end{equation}
we have
\begin{subequations}\label{e:VV}
\begin{align}
   V_1(x)
   &= 
   - ( \, b^2 + m^2 - 1/4 ) \, \sech^2(x) \>,
   \label{e:VVa} \\
   V_2(x)
   &= 
   - 2 \, m b  \tanh(x) \sech(x) \>.
   \label{e:VVb}
\end{align}
\end{subequations}
Note this potential is invariant under the exchange of $b$ and $m$. 
We are interested in the stability properties of the exact solutions of the NLS equation in this external potential:
\begin{equation}\label{e:NLS equation-1}
  \{
   i \,
   \partial_t
   +  
   \partial_x^2
   +
   g  | \psi(x,t) |^{2 \kappa}
   - 
   [\, V_1(x) + i V_2(x) \,] \,
   \} \, \psi(x,t) 
   = 
   0 \>.
\end{equation}
This equation can be obtained from a generalized Euler-Lagrange equation using a dissipation functional \cite{PhysRevE.94.032213},
\begin{equation}\label{e:euler-lagrange}
   \frac{\delta \Gamma}{\delta \psi^{\ast}}
   =
   - \frac{\delta \mc{F}}{\delta \psi_t^{\ast}} \>,
\end{equation}
where
\begin{subequations}\label{e:defGF}
\begin{align}
   \Gamma
   &=
   \tint \dd{t}
   \Bigl \{ \, 
      \frac{\rmi}{2} \tint \dd{x}
      [\, \psi^{\ast} \psi_t - \psi \psi_t^{\ast} \,]
      -
      H \,
   \Bigr \} \>,
   \label{e:defGF-a} \\
   H
   &=
   \tint \dd{x}
   \Bigl \{ \,
      | \psi_x |^2
      -
      \frac{g \, | \psi |^{2\kappa +2}}{\kappa + 1} 
      +
      V_1(x) \, | \psi |^2 \,
   \Bigr \} \>,
   \label{e:defGF-b} \\
   \mc{F}
   &=
   \tint \dd{t} F
   =
   \rmi \tint \dd{x} \dd{t} V_2(x) \, 
   [\, \psi_t \, \psi^{\ast} - \psi_t^{\ast} \, \psi \,] \>.
   \label{e:defGF-c}
\end{align}
\end{subequations}
Localized solutions to Eq.~\ef{e:NLS equation-1} exist for arbitrary values of $\kappa, m, b$.  Here we use $\psi_0(x,t)$ to denote the exact solution to the NLS equation in the external potential,
\begin{equation}\label{psi0wf}
   \psi_0(x,t) 
   = 
   A_0 \, \sech^{1/\kappa}(x)\, \rme^{\rmi [\, Et + \phi(x)\,] } \, 
\end{equation}
where 
\begin{equation}
    \phi(x) 
    = 
    \frac{4 b m \kappa}{\kappa + 2} \,
    \tan^{-1}[\tanh(x/2)\,] \>,
\end{equation}
with $E = 1/\kappa^2$, and
\begin{equation}\label{gasq}
   g A_0^{2 \kappa} 
   = 
   \frac{[4 b^2 \kappa^2-(\kappa + 2)^2]\, [4 m^2 \kappa^2 -(\kappa+2)^2]}
        {4 \kappa^2 (\kappa+2)^2} \>.
\end{equation}
We notice when $mb=0$ the potential is real and that solutions exist for $m^2+ b^2-1/4 < (\kappa+1)/\kappa^2$.  There are two regimes where $A_0^2$ is positive and so a solution exists when $m \neq 0$.  This form of the solution reflects the fact that the potential $V^+$ is invariant under the interchange $m \leftrightarrow b$.
 
In a previous paper \cite{r:Dawson:2017td} we studied the stability of these solutions for $m=0$ (real external potential) and for arbitrary $\kappa$.  In that paper, we also considered two other cases where exact solutions exist. For the case of $g = -1$ and attractive potential, for $V_2=0$, all the solutions that were allowed were stable.
Solutions also exist for $V_2 \neq 0$ and are  given by Eqs.~\ef{psi0wf} and \ef{gasq} with  $g =-1$. For $g=1$ and a repulsive real potential we found the solutions for $V_2=0$ were translationally unstable.  Solutions again exist when $V_2 \neq 0$ for this case.  We will not discuss these solutions further here. 

Here we will confine ourselves to $\kappa=1, g=1$ and an attractive external potential $V_1$  and study the domain of applicability of the variational methods we have developed previously to the case of increasing the dissipation by allowing $m$ to vary.   In particular for the case we will concentrate on here  ($\kappa=1$) we have that 
\begin{equation}\label{gAsq}
   g A_0^{2} = (4 b^2 -9)(4m^2 -9) / 36 \>,
\end{equation}
so that when $m^2 < 9/4$ we need that $b^2 < 9/4$ for there to be a solution.  Also if we confine ourselves to an attractive potential so that we avoid the known translational instability associated with repulsive potentials \cite{r:Dawson:2017td}, then we also require $b^2 + m^2 > 1/4$.  Note that $g A^2_0$ is independent of $g$.  For $\kappa=1$ we have
\begin{subequations}\label{e:NLS equation-4}
\begin{align}
   \phi(x)
   &=
   ( 4 \, m \, b/3 ) \, \tan^{-1}[\, \tanh(x/2) \,] \>,
   \label{e:NLS equation-4a} \\
   \partial_x \phi(x)
   &=
   (2/3) \, m \, b \, \sech(x) \>.
   \label{e:NLS equation-4b}
\end{align}
\end{subequations}

%
%
\section{\label{s:general}Some General Properties of the NLS equation in complex potentials}
 
We are interested in  solitary wave solutions that approach zero exponentially  at $\pm \infty$.  For these solutions we define the mass density $\rho(x,t) = \abs{\psi(x,t)}^2$, and 
the mass or norm  $M(t)$  as
\begin{equation}\label{eqMass1}
   M(t) 
   =
   \tint \dd{x} \rho(x,t)  
   = 
   \tint \dd{x} \abs{\psi(x,t)}^2 \>.
\end{equation}
In addition, we define the current as: 
\begin{equation}
   j(x,t) 
   = i \,
   [\, 
      \psi(x,t) \, \psi_x^\ast(x,t) 
      - 
      \psi^{\ast}(x,t) \, \psi_x(x,t) \,
   ] \>.
\end{equation}
Multiplying the NLS equation \ef{e:NLS equation-1} by $\psi^{\ast}(x,t)$ and subtracting the complex conjugate of the resulting equation, we obtain
\begin{equation}\label{cont}
   \pdv{\rho(x,t)}{t} 
   +  
   \pdv{j(x,t)}{x}  
   =  
   2 V_2(x) \, \rho(x,t).
\end{equation}
Integrating over space, and assuming that $j(+\infty,t) - j(-\infty,t) = 0$, we find 
\begin{equation}\label{mdot}
   \dv{M(t)}{t}
   =  
   2 \tint \dd{x} V_2(x) \, \rho (x,t) \>. 
\end{equation}
Note that $M$ is conserved when $V_2(x)=0$.   
If we instead multiply the NLS equation by $\psi^\ast$ and add the complex conjugate of the resulting equation, we get
\begin{align} \label{add}
   & \rmi \, (\,  \psi^\ast \psi_t - \psi \psi^\ast_t \,) 
   \\
   & \qquad 
   = 
   - 
   2 g \rho^2 
   - 
   \psi^\ast \psi_{xx} 
   - 
   \psi \psi^\ast_{xx} 
   + 
   2 V_1(x) \, \rho \>,
   \notag
\end{align}
which when we integrate over space, leads to the virial theorem:
\begin{align}
   & \frac{\rmi}{2} 
   \tint \dd{x} 
   (\,  \psi^{\ast} \psi_{t} - \psi_t^\ast \psi \,) 
   - 
   \tint \dd{x}  
   \qty [\,
      \abs{\psi_x}^2 
      - 
      g \, \abs{\psi}^{4} \,
        ]
   \\
   & \qquad\qquad
   = 
   \tint \dd{x} V_1(x) \, \abs{\psi}^2 \>.
   \notag
\end{align}
The average position $q(t)$ can be defined through the first moment of $x$ as follows:
\begin{equation}
   M_1(t) 
   = 
   \tint \dd{x} x \, \rho(x,t) = q(t) M(t) \>.
\end{equation}
Multiplying the continuity equation \eqref{cont} by $x$ and integrating over all space we find:
\begin{equation}
   \dv{M_1}{t}  
   = 
   2 \, P(t) 
   + 
   2 \tint \dd{x} \,
   x \,  V_2(x) \, \rho(x,t) \>,
\end{equation}
where the momentum 
\begin{align}
   P(t) 
   &=  
   \frac{1}{2} \tint \dd{x} j(x,t)
   \\
   &=
   \frac{\rmi}{2}
   \tint \dd{x} 
   \qty[\, 
      \psi^\ast(x,t) \, \psi_x(x,t) 
      - 
      \psi^{\ast}_x(x,t) \psi(x,t) \,] \>.
   \notag
\end{align}
Here, we assumed that
\begin{equation}\label{e:assumedlimits}
   \lim_{x\rightarrow \infty} x j(x,t) 
   - 
   \lim_{x\rightarrow -\infty} x j(x,t) 
   = 0 \>.   
\end{equation}
Assuming that the density is a function of $y = x - q(t)$ and $t$, we find
\begin{align*}
   \dv{t} \, \qty[\, M(t) \,q(t) \,] 
   &=
   2 \,  P(t)
   + 
   2 \tint \dd{y} y  \, V_2(y+q(t))  \, \rho(y,t) 
   \\
   & \qquad
   +
   2 \, q(t) \tint \dd{x} \, V_2(x) \, \rho(x-q(t),t) \>.
\end{align*}
We recognize the last term as $q(t) \dd{M(t)}/\dd{t}$, so that we finally have:
\begin{equation}\label{dotq1}
   M(t) \, 
   \dv{q(t)}{t} 
   = 
   2 \, P(t) 
   + 
   2 \tint \dd{y} y \, V_2(y+q(t)) \, \rho(y,t) \>.
\end{equation}
Taking the time derivative of the momentum $P (t)$, using the equations of motion for $\psi$ and $\psi^{\ast}$, and integrating by parts, we find
\begin{equation}\label{pdot}
   \dv{P(t)}{t}  
   = 
   - 
   \tint \dd{x} \rho (x,t) \pdv{V_1(x)}{x}  
   +  
   \tint \dd{x} j(x,t ) \, V_2(x) \>. 
\end{equation}
Here 
\begin{equation}
   \pdv{V_1(x)}{x} 
   =
   2 \, \qty( \, b^2 + m^2 - 1/4 \,) \, 
   \tanh(x) \, \sech^2(x) \>.
\end{equation}
Note that in our case $V_1(x) $ is an even function of $x$ and $V_2(x)$ is  an odd function.  In our study we will assume $\rho(x,t) = \tilde \rho (y, t)$ where $y(t) = x-q(t)$.  That is, the functional form of $\rho$ will be maintained if it is given a slight perturbation away from the origin.  If it stays at the origin ($q(t) =0$) and only changes its width and amplitude under perturbation, then we see that since $\rho$ is an even function of $y$ and $V_2(x)$ is an odd function of $x$, the mass is conserved.  One can in a systematic fashion obtain the equations for the higher moments of $ \langle x^n \, \hat{p}^m \rangle $, where $\hat{p} = -i \partial/\partial x$.  It can be demonstrated that the four and six collective coordinate approximations we derive in this paper will satisfy a particular subset of four or six moment equations \cite{PhysRevE.82.016606}.

%
%
\section{\label{s:CCs}Collective coordinates}

The time dependent variational approximation relies on 
 introducing  a finite set of time-dependent real parameters in a  trial wave function that one hopes captures the time evolution of a perturbed solution.  By doing this   one obtains a simplified set of ordinary differential equations for the collective coordinates in place of solving the full partial differential equation for the NLS equation. By judiciously choosing the collective coordinates, they can be simply related to the moments of  $x$ and ${\hat p}= -i \partial/\partial x $ averaged over the density $\rho(x,t)$. 
 
That is, we set
\begin{align}\label{e:VT-1}
   \psi(x,t)
   &\mapsto
   \tilde{\psi}[\,x,Q(t)\,] 
   \\
   Q(t) 
   &= 
   \{\, Q^1(t),Q^2(t),\dotsc,Q^{2n}(t) \,\} \in \mathbb{R}^{2n} \>.
   \notag
\end{align}
The success of the method depends greatly on the choice of the the trial wave function $\tilde{\psi}[\,x,Q(t)\,]$.  The generalized Euler-Lagrange equations lead to Hamilton's equations for the collective coordinates $Q(t)$.
Introducing the notation $\partial_{\mu} \equiv \partial / \partial Q^{\mu}$, the Lagrangian in terms of the collective coordinates is given by
\begin{equation}\label{e:VT-2}
   L(\,Q,\dot{Q}\,)
   =
   \pi_\mu(Q) \, \dot{Q}^\mu - H(\,Q\,) \>,
\end{equation}
where $\pi_\mu(Q)$ is defined by
\begin{align}\label{e:VT-3}
   \pi_\mu(Q)
   &=
   \frac{\rmi}{2} \tint \dd{x}
   \{ \, 
      \tpsi^{\ast}(x,Q)\,[\, \partial_\mu \tpsi(x,Q) \,]
      \\
      & \qquad\qquad
      - 
      [\, \partial_\mu \tpsi^{\ast}(x,Q) \,] \, \tpsi(x,Q)
   \,\} \>,
   \notag
\end{align}
and $H(Q)$ is given by
\begin{align}\label{e:VT-4}
   H(Q)
   &=
   \tint \dd{x} 
   \Bigl \{ \,
       |\partial_x \tpsi(x,Q) |^2
       -
       \frac{g}{2} \, |\tpsi(x,Q)|^{4} 
       \\
       & \qquad\qquad
       +
       V_1(x) \, |\tpsi(x,Q)|^2 \,
    \Bigr \} \>.
    \notag   
\end{align}
Similarly, in terms of the collective coordinates, the dissipation functional is given by
\begin{equation}\label{e:VT-4.1}
   F[Q,\dot{Q}]
   =
   w_{\mu}(Q) \, \dot{Q}^{\mu} \>,
\end{equation}
where
\begin{align}\label{e:VT-4.2}
   w_{\mu}(Q)
   &=
   \rmi \tint \dd{x} V_2(x) \,
   \{ \, 
      \tpsi^{\ast}(x,Q)\,[\, \partial_\mu \tpsi(x,Q) \,]
      \\
      & \qquad\qquad
      - 
      [\, \partial_\mu \tpsi^{\ast}(x,Q) \,] \, \tpsi(x,Q)
   \,\} \>.
   \notag
\end{align}
The generalized Euler-Lagrange equations are
\begin{equation}\label{e:VT-5}
   \pdv{L}{Q^\mu}
   -
   \dv{t} \Bigl ( \pdv{L}{\dot{Q}^\mu} \Bigr )
   =
   -
   \pdv{F}{\dot{Q}^\mu} \>.
\end{equation}
Setting $v_{\mu}(Q) = \partial_\mu H(Q)$, we find
\begin{equation}\label{e:VT-6}
   f_{\mu\nu}(Q) \, \dot{Q}^\nu
   =
   u_{\mu}(Q)
   =
   v_{\mu}(Q) - w_{\mu}(Q) \, 
\end{equation}
where 
\begin{equation}\label{e:VT-7}
   f_{\mu\nu}(Q)
   =
   \partial_\mu \pi_\nu(Q) - \partial_\nu \pi_\mu(Q)
\end{equation}
is an antisymmetric $2n \times 2n$ symplectic matrix.  
If $\det{f(Q)} \ne 0$, we can define an inverse as the contra-variant matrix with upper indices,
\begin{equation}\label{e:VT-8}
   f^{\mu\nu}(Q) \, f_{\nu\sigma}(Q) = \delta^\mu_\sigma \>,
\end{equation}
in which case the equations of motion \ef{e:VT-6} can be put in the symplectic form:
\begin{equation}\label{e:VT-9}
   \dot{Q}^\mu
   =
   f^{\mu\nu}(Q) \, u_{\nu}(Q) \>.
\end{equation}
Poisson brackets are defined using $f^{\mu\nu}(Q)$.  If $A(Q)$ and $B(Q)$ are functions of $Q$, Poisson brackets are defined by
\begin{equation}\label{e:PB-1}
   \PB{A(Q)}{B(Q)}
   =
   ( \partial_\mu A(Q) ) \, f^{\mu\nu}(Q) \, ( \partial_\nu B(Q) ) \>.
\end{equation}
In particular,
\begin{equation}\label{e:PB-2}
   \PB{Q^\mu}{Q^\nu}
   =
   f^{\mu\nu}(Q) \>.
\end{equation}
It is easy to show that $f_{\mu\nu}(x)$ satisfies Bianchi's identity.  This means that definition \ef{e:PB-1} satisfies Jacobi's identity, as required for symplectic variables.  The rate of energy loss is expressed as
\begin{equation}\label{e:EC-1}
   \dv{H(Q)}{t}
   =
   - v_\mu(Q) \, f^{\mu\nu}(Q) \, w_{\nu}(Q) \>,
   \notag
\end{equation}
since $f^{\mu\nu}(Q)$ is an antisymmetric tensor.

%
%
\section{\label{s:FourPar}Four parameter trial wave function}

Let us first look at the four parameter trial wave function that we have successfully used to study the effect of weak complex external potentials on the exact solution of the NLS equation in the absence of that potential.  That is we will choose:
\begin{equation}\label{e:T4-1}
   \tpsi(x,t)
   =
   A_0 \beta(t) \, \sech[\, \beta(t) \, y(x,t) \,] \, 
   \rme^{\rmi \, \tphi(x,t)} \>,
\end{equation}
where $A_0$ is the amplitude of the exact solution in the presence of the external potential \ef{gAsq} and is a funcition of  $m,b,g$, and  
\begin{equation}\label{e:T4-2}
   \tphi(x,t)
   =
   - 
   \theta(t)
   +
   p(t) \, y(x,t)
   +
   \phi(x) \>.
\end{equation}
Here $\phi(x)$ is given by Eq.~\ef{e:NLS equation-4} and we have put $y(x,t) = x - q(t)$.  
The four variational parameters are labeled by
\begin{equation}\label{e:T-2.1}
   Q^{\mu}
   =
   \qty{\, q(t), p(t), \beta(t) , \theta(t) \,} \>. 
\end{equation}
The  derivatives of $\tpsi(x,t)$ with respect to $t$ and $x$ are given by
\begin{subequations}\label{e:T-3}
\begin{align}\label{e:T-3a}
   &\tpsi_t(x,t)
   =
   A_0 \, 
   \{\,
      \dot{\beta} \sech( \beta y )
      \\
      & \qquad\qquad
      -
      \beta \sech(\beta y) \, \tanh(\beta y) \,
      [\, \dot{\beta} y - \dot{q} \beta \,]
    \notag\\
         & \qquad
      {}+
      \rmi \, \beta \sech( \beta \, y \,) \,
      [\, - \dot{\theta} + \dot{p} \, y - p \, \dot{q} \,] \, 
   \} \, \rme^{\rmi \, \tphi(x,t)} \>,
   \notag \\
   &\tpsi_x(x,t)
   =
   A_0 \, \beta \,
   \{\,
      -
      \beta \, \sech(\beta y) \, \tanh(\beta y) \,
      \label{e:T-3b} \\
      & \qquad
      +
      \rmi \, \sech(\beta y) \,
      [\, p + (2/3) \, m \, b \, \sech(x) \,] \,
   \} \, \rme^{\rmi \, \tphi(x,t)} \>, 
   \notag
\end{align}
\end{subequations}
where we have used \ef{e:NLS equation-4b}.  Then the density and current is given by
\begin{subequations}\label{e:T-4}
\begin{align}
   \rho(x,t)
   &=
   A_0^2 \, \beta^2 \sech^2(\beta y) \>,
   \label{e:T-6a} \\
   j(x,t)
   &=
   2 \, \rho(x,t) \,
      [\,
         p
         +
         (2/3) \, m \,  b \sech(x)\,
      ] \>.
   \label{e:T-6b}
\end{align}
\end{subequations}
The time dependent mass, $M(t)$ which is a normalization factor, is given by
\begin{equation}\label{e:T-4.1}
   M(t)
   =
   \tint \dd{x} \rho(x,t)
   =
   2 \, A_0^2 \, \beta(t) \>, 
\end{equation}
and the Lagrangian and dissipation function are given by,
\begin{subequations}\label{e:T-7}
\begin{align}
   L
   &=
   \frac{\rmi}{2} \tint \dd{x}
   \qty[\, \psi^{\ast}\,\psi_t - \psi^{\ast}_t\,\psi \,]
   -
   H[\,\psi,\psi^{\ast}\,] \>,
   \label{e:T-7a} \\
   H
   &=
   \tint \dd{x}
   [\,
      \abs{ \psi_x }^2
      -
      g \, \abs{ \psi }^4 / 2
      +
      V_1(x) \, \abs{ \psi }^2 \,] \>,
   \label{e:T-7b} \\
   F
   &=
   \rmi \tint \dd{x} V_2(x) \,
   \qty[\, \psi^{\ast} \, \psi_t - \psi^{\ast}_t \, \psi \,] \>.
   \label{e:T-7c}
\end{align}
\end{subequations}
The generalized Euler-Lagrange equations are
\begin{subequations}\label{e:T-8}
\begin{align}
   \fdv{L}{\psi^{\ast}}
   -
   \partial_t \fdv{L}{\psi_t^{\ast}}
   &=
   - \fdv{F}{\psi_t^{\ast}} \>,
   \label{e:Av-3a} \\
   \fdv{L}{\psi^{\phantom\ast}}
   -
   \partial_t \fdv{L}{\psi_t^{\phantom\ast}}
   &=
   - \fdv{F}{\psi_t^{\phantom\ast}} \>. 
   \label{e:Av-3b} 
\end{align}
\end{subequations}
For the trial wave function of Eq.~\ef{e:T4-1}, we find
\begin{align}\label{e:T-9}
   L_0[Q]
   &\equiv
   \frac{\rmi}{2} \tint \dd{x}
   [\, \tpsi^{\ast}\,\tpsi_t - \tpsi^{\ast}_t\,\tpsi \,]
   \\
   &=
   2 A_0^2 \, \beta \, 
   (\, \dot{\theta} + p \, \dot{q} \,)
   \equiv
   \pi_\mu(Q) \, \dot{Q}^\mu \>,
   \notag
\end{align}
where
\begin{equation}\label{e:T-10}
   \pi_{q} = 2 A_0^2 \, \beta \, p
   \qc
   \pi_{p} = 0 
   \qc
   \pi_{\beta} = 0
   \qc
   \pi_{\theta} = 2 A_0^2 \, \beta \>.
\end{equation}
The only partial derivatives of $\pi_\mu(Q)$ that survive are:
\begin{equation}\label{e:T-11}
   \partial_p \pi_{q} = 2 A_0^2 \, \beta
   \qc
   \partial_{\beta} \pi_{q} = 2 A_0^2 \, p
   \qc
   \partial_{\beta} \pi_{\theta} = 2 A_0^2 \>.
\end{equation}
So the symplectic matrix and its inverse are given by
\begin{align}\label{e:T-12}
   f_{\mu\nu}(Q)
   &=
   2 A_0^2
   \begin{pmatrix}
      0 & -\beta & -p & 0 \\
      \beta & 0 & 0 & 0 \\
      p & 0 & 0 & 1 \\
      0 & 0 & -1 & 0
   \end{pmatrix} \>,
   \\
   f^{\mu\nu}(Q)
   &=
   \frac{1}{2 A_0^2 \, \beta}
   \begin{pmatrix}
      0 & 1 & 0 & 0 \\
      -1 & 0 & 0 & p \\
      0 & 0 & 0 & -\beta \\
      0 & -p & \beta & 0
   \end{pmatrix} \>.
   \notag
\end{align}
From the Hamiltonian \ef{e:T-7b} and our choice of trial wave function we find that 
\begin{align}
   &H(Q)
   =
   A_0^2 \, \beta \,
   \{\,
      (2/3) \, \beta^2
      + 
      2 \, p^2
      +
      (4/3) \, p \, m \, b \, \beta \, I_1(\beta,q) 
      \notag \\
      &
      - 
      [\, b^2 + m^2 - (4/9) \, m^2 b^2 - 1/4 \,] \, \beta \, I_2(\beta,q) 
   \,\}
   \label{e:T-16} \\
   & \qquad\qquad
   -
   (2/3) \, g \, A_0^4 \, \beta^3 \>, 
   \notag
\end{align}
where $I_1(\beta,q)$ and $I_2(\beta,q)$ are given in Appendix~\ref{s:integrals}.
Then defining $v_{\mu} = \partial_{\mu} H(Q)$, we find 
\begin{subequations}\label{e:T-18}
\begin{align}
   v_q
   &=
   - A_0^2 \, \beta \,
   [\,
      (4/3) \, p \, m \, b \, \beta \, f_1(\beta,q)
      \label{e:T-18a} \\
      & \qquad
      -
      2 \, [\, b^2 + m^2 - (4/9) \, m^2 b^2 - 1/4 \,] \, 
      \beta \,  f_{6}(\beta,q) \,
   ] \>,
   \notag \\
   v_p
   &=
   A_0^2 \, \beta \,
   [\, 4 p + (4/3) \, m \, b \, \beta \, I_1(\beta,q) \,] \>,
   \label{e:T-18b} \\
   v_\beta
   &=
   A_0^2 \, \beta \,
   \{\,
      2 \, \beta
      +
      2 \, p^2 / \beta
      \label{e:T-18c} \\
      & \qquad
      +
      (8/3) \, p \, m \,  b \, 
      [\, I_1(\beta,q) - \beta \, f_{10}(\beta,q) \,]
      \notag \\
      & \qquad
      -
      2 \, [\, b^2 + m^2 - (4/9) \, m^2 b^2 - 1/4 \,]
      \notag \\
      & \qquad
      \times
      [\, I_2(\beta,q) - \beta \, f_7(\beta,q) \,]
      -
      2 \, g \, A_0^2 \, \beta \,
   \} \, 
   \notag \\
   v_\theta
   &=
   0 \>,
   \label{e:T-18d} 
\end{align}
\end{subequations}
where the $f_i(\beta,q)$ are given in Appendix~\ref{s:integrals}.
From \ef{e:T-7c}, the dissipation function is given by
\begin{align}\label{e:T-19} 
   F[Q,\dot{Q}]
   =
   w_\mu(Q) \, \dot{Q}^\mu \>,
\end{align}
where
\begin{subequations}\label{e:T-20}
\begin{align}
   w_q
   &=
   - 4 \, m \, b \, A_0^2 \, \beta^2 \, p \, f_1(\beta,q) \>,
   \label{e:T-20a} \\
   w_p
   &=
   4 \, m \, b \, A_0^2 \, \beta^2 \, f_2(\beta,q) \>,
   \label{e:T-20b} \\
   w_\beta
   &=
   0 \>,
   \label{e:T-20c} \\
   w_\theta
   &=
   - 4 \, m \, b \, A_0^2 \, \beta^2 \, f_1(\beta,q) \>.
   \label{e:T-20d}
\end{align}
\end{subequations}
Here $f_1(\beta,q)$ and $f_2(\beta,q)$ are given in Appendix~\ref{s:integrals}.
In terms of the vector $u_{\mu}(Q) = v_{\mu}(Q) - w_{\mu}(Q)$,
Hamilton's equations for the variational parameters are
\begin{equation}\label{e:T-22}
   \dot{Q}^\mu
   = 
   f^{\mu\nu}(Q) \, u_\nu(Q) \>, 
\end{equation}
which gives
\begin{subequations}\label{e:T-24}
\begin{align}
   \dot{q}
   &=
   2 \, p 
   + 
   (2/3) \, m \, b \, \beta \, I_1(\beta,q) 
   - 
   2 \, m \, b \, \beta \, f_2(\beta,q) \>,
   \label{e:T-24a} \\
   \dot{p}
   &=
   (2/3) \, p \, m \, b \, \beta \, f_1(\beta,q)
   \label{e:T-24b} \\
   & \qquad
   -
   [\, b^2 + m^2 - (4/9) \, m^2 b^2 - 1/4 \,] \, \beta \, f_6(\beta,q) \, 
   \notag \\
   \dot{\beta}
   &=
   - 2 \, \beta^2 \, m \, b \, f_1(\beta,q) \>.
   \label{e:T-24c}
\end{align}
\end{subequations}
The equation for $\dot{\theta}$ is not needed for the evolution of the set of equations given in \ef{e:T-24}.  For $m=0$, the equations reduce to:
\begin{subequations}\label{e:T-25}
\begin{align}
   \dot{q}
   &=
   2 \, p \>,
   \label{e:T-25a} \\
   \dot{p}
   &=
   -
   [\, b^2 -1/4 \,] \, \beta \, f_6(\beta,q) \>,
   \label{e:T-25b} \\
   \dot{\beta}
   &=
   0 \>.
   \label{e:T-25c}
\end{align}
\end{subequations}
So in this case, $\beta = 1$ and is fixed.  This is because the normalization must be conserved.  Equations~\ef{e:T-25} then reduce to:
\begin{equation}\label{e:T-26}
   \ddot{q}
   +
   2 \, [\, b^2 - 1/4 \,] \, f_6(1,q)
   =
   0 \>.
\end{equation}

%
%
\subsection{\label{s:smallosc}Small Oscillation equations}  

\begin{figure}[t]
   \centering
   \includegraphics[width=0.9 \columnwidth]{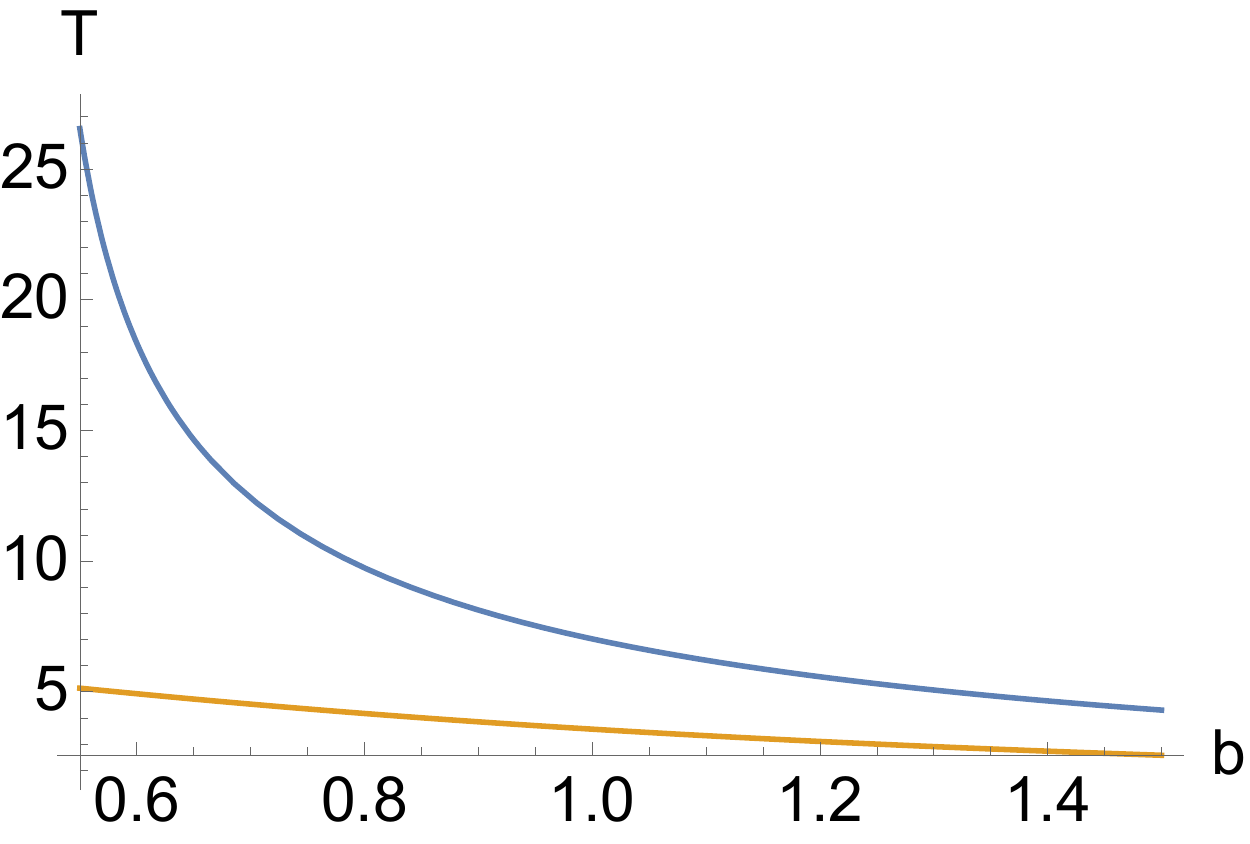}
   \caption{\label{f:fig1} Period as a function of $b$ for $m=0$ 
   (upper curve) and $m=1$ (lower curve)  for 4 CC approximation.}
\end{figure}

Using the expansions found in Appendix A we obtain for the small oscillation equations (we set $q = \delta q$, $p = \delta p$, and $\beta = 1 + \delta \beta$ with $\delta Q^{\mu}$ assumed small),
\begin{subequations}\label{e:SO4-1}
\begin{align}
   \delta \dot{q}
   & =
   \frac{\pi}{72} \,
   \qty( 9 \pi^2 - 64 ) b m \, \delta \beta + 2 \, \delta p \>,
   \label{e:SO4-1a} \\
   \delta \dot{p}
   &=
   -   
   \frac{8}{15}
   \qty( \,
      b^2
      +
      m^2
      -
      (4/9) \, b^2 m^2
      -
      1/4 \, ) \, \delta q \>,
   \label{e:SO4-1b} \\
   \delta \dot{\beta}
   &=
   - \frac{\pi}{2} \, m b \, \delta q \>.
   \label{e:SO4-1c}
\end{align}
\end{subequations}
Thus we obtain for  $\ddot{q}$ 
\begin{equation}
   \delta\ddot{q} 
   + 
   \omega^2(b,m) \, \delta q  
   = 
   0 \>,
\end{equation}
where
\begin{align}
   \omega^2(b,m)
   &=
   \frac{\pi^2}{144} \, 
   (\, 9 \pi^2 - 64 \, ) \, b^2 m^2
   \\
   & \qquad
   +
   \frac{16}{15} \,
   \qty(
      b^2
      +
      m^2
      -
      (4/9) \, b^2 m^2
      -
      1/4 \,) \>. 
      \notag
\end{align}
The period $T = 2 \pi/ \omega(b,m)$ for $m=0$ and $m=1$ is shown in Fig.~\ref{f:fig1}.
%
%
%
%

%
%
\section{\label{s:6cc}Six parameter ansatz}

One expects that when one increases the number of CC's  the accuracy of the variational approximation increases.  For the six parameter Ansatz we will introduce a ``chirp'' term \cite{PhysRevD.34.3831} $\Lambda(t)$ which is conjugate to the width parameter $\beta(t)$.   That is we will assume:
\begin{equation}\label{e:T-1}
   \tpsi(x,t)
   =
   A(t)  \, \sech[\, \beta(t) \, y(x,t) \,] \>, 
   \rme^{\rmi \, \tphi(x,t)} \,
\end{equation}
where
\begin{equation}\label{e:T-2}
   \tphi(x,t)
   =
   - 
   \theta(t)
   +
   p(t) \, y(x,t)+ \Lambda(t) y(x,t) ^2
   +
   \phi(x) \>.
\end{equation}
Here $\phi(x)$ is given by Eq.~\ef{e:NLS equation-4} and we have put $y(x,t) = x - q(t)$.  We find  
\begin{equation}\label{e:P-3}
   \rho(x,t)
   =
   | \tpsi(x,t) |^2
   =
   A^2(t) \sech^2(\beta y) \>,
\end{equation}
so that the mass becomes
\begin{equation}\label{e:P-4}
   M(t)
   =
   \tintx \rho(x,t)
   =
   \frac{2 A^2(t)}{\beta(t)} \>.
\end{equation}
It will be useful to employ $M(t)$ as a collective coordinate rather than $A(t)$.  The six time-dependent collective coordinates then are:
\begin{equation}\label{e:P-5}
   Q^{\mu}(t)
   =
   \{\, M(t), \theta(t), q(t), p(t), \beta(t), \Lambda(t) \,\} \>.
\end{equation}

The parameters $\beta(t)$ and $\Lambda(t)$ are related to the two point correlation functions   $G_2 = \langle (x-q(t))^2 \rangle$ and 
$P_2 = \ev{[x-q(t) ] \hat p + \hat p [x-q(t)] }$ where 
\begin{equation}\label{e:evdef}
   \ev{ (\cdot) } 
   =
   \int\limits_{-\infty}^{\infty} (\cdot)|\psi(x,t)|^2 \dd{x}
   \Big /
   \int\limits_{-\infty}^{\infty}|\psi(x,t)|^2 \dd{x} \>.
\end{equation}
Thus we find $G_2 = \pi^2 / (12 \beta^2)$, and
\begin{align}\label{e:P2} 
   P_2 
   &= 
   \frac{\rmi}{2} \tint \dd{x}
   [x-q(t)] \,
   \qty[\, \psi^\ast \psi_x - \psi^\ast_x \psi \,] / M(t)
   \\
   &= 
   \frac{\pi^2 \Lambda}{3 \beta^2}  
   + 
   \frac{2}{3} b m  \frac{I_3(\beta,q) }{M(t)} \>,
   \notag
\end{align}
where $I_3$ is given in Appendix~\ref{s:integrals}. We see that $P_2$ is directly related to $\Lambda$ when the potential is real.

From the formalism given in Sec.~\ref{s:CCs}, the equations of motion for the collective coordinates follow. 
For the kinetic term in the Lagrangian, we find
\begin{gather}\label{e:P-12}
   \pi_{M} = 0
   \qc
   \pi_{\theta} = M
   \qc
   \pi_{q} = M p
   \qc
   \pi_{p} = 0 \, 
   \\
   \pi_{\beta} = 0 
   \qc
   \pi_{\Lambda} = - M \frac{\pi^2}{12 \beta^2} \>,
   \notag
\end{gather}
and the only non-zero derivatives are then
\begin{gather}\label{e:P-13}
   \partial_M \pi_{\theta}
   =
   1
   \qc
   \partial_M \pi_{q}
   =
   p
   \qc
   \partial_p \pi_{q}
   =
   M \, 
   \\
   \partial_M \pi_{\Lambda}
   =
   - \frac{\pi^2}{12 \beta^2}
   \qc
   \partial_{\beta} \pi_{\Lambda}
   =
   M \frac{\pi^2}{6 \beta^3} \>.
   \notag
\end{gather}
\begin{widetext}
The antisymmetric symplectic tensor is then given by
\begin{equation}\label{e:P-15}
   f_{\mu\nu}(Q)
   =
   \begin{pmatrix}
      0 & 1 & p & 0 & 0 & -\pi^2/(12 \beta^2) \\
      -1 & 0 & 0 & 0 & 0 & 0 \\
      - p & 0 & 0 & -M & 0 & 0 \\
      0 & 0 & M & 0 & 0 & 0 \\
      0 & 0 & 0 & 0 & 0 & M \pi^2/(6 \beta^3) \\
      \pi^2/(12 \beta^2) & 0 & 0 & 0 & -M \pi^2/(6 \beta^3) & 0
   \end{pmatrix} \>.
\end{equation}
Since $\det{f_{ij}(Q)} = M^4 \pi^4/( 36 \beta^6)$ and is non-zero, the inverse is given by
\begin{equation}\label{e:P-16}
   f^{\mu\nu}(Q)
   =
   \begin{pmatrix}
      0 & -1 & 0 & 0 & 0 & 0 \\
      1 & 0 & 0 & -p/M & \beta/(2M) & 0 \\
      0 & 0 & 0 & 1/M & 0 & 0 \\
      0 & p/M & -1/M & 0 & 0 & 0 \\
      0 & -\beta/(2M) & 0 & 0 & 0 & -6 \beta^3/(\pi^2 M) \\
      0 & 0 & 0 & 0 & 6 \beta^3/(\pi^2 M) & 0
   \end{pmatrix} \>.
\end{equation}
For the dissipation functional, we obtain
\begin{align}\label{e:P-26}
   F(Q,\dot{Q})
   &=
   2 M m b \beta
   \tinty \sech^2( \beta y ) \sech(y+q) \tanh(y+q)  \,
   [\,
      -
      \dot{\theta}
      + 
      \dot{p} y
      -
      p \dot{q}
      +
      \dot{\Lambda} y^2 
      -
      2 y \Lambda \dot{q} \,
   ] \>,
\end{align}
which gives
\begin{gather}\label{e:P-27}
   w_{M} = 0
   \qc
   w_{\theta}
   =
   - 2 M m b \beta \, f_1(\beta,q)
   \qc
   w_{q}
   =
   - 2 M \, m b \beta \,
   [\, p \, f_1(\beta,q) + 2 \Lambda f_2(\beta,q) \,]
   \qc \\
   w_{p}
   =
   2 M \, m b \beta \, f_2(\beta,q)
   \qc
   w_{\beta}
   =
   0
   \qc
   w_{\Lambda}
   =
   2 M \,  m b \beta \, f_3(\beta,q) \>.
   \notag
\end{gather}
For $H(Q)$, using the 6-parameter Ansatz we now obtain
\begin{align}\label{e:P-23}
   \frac{H(Q)}{M} 
   &=
   p^2
   +
   \frac{\beta^2}{3} 
   +
   \frac{ \pi^2 \Lambda^2}{3 \beta^2}
   +
   \frac{2 \beta}{3} \, p b m \, I_1(\beta,q)
   +
   \frac{4 \beta}{3} \, b m \Lambda \, I_3(\beta,q)
   \\
   & \qquad\qquad
   - 
   \frac{g M \beta}{6}
   - 
   \frac{\beta}{2} \, 
   \Bigl [\,
       b^2 + m^2 - \frac{1}{4} 
       - \frac{4}{9}\, b^2 m^2 \,
   \Bigr ] \, I_2(\beta,q) \>.
   \notag
\end{align}
All the integrals are defined in Appendix~\ref{s:integrals}.
For  $v_{\mu}(Q) = \partial_{\mu} H(Q)$ we obtain
\begin{subequations}\label{e:P-24}
\begin{align}
   v_{M}
   &=
   p^2
   +
   \frac{\beta^2}{3} 
   +
   \frac{ \pi^2 \Lambda^2}{3 \beta^2}
   +
   \frac{2 \beta}{3} \, p b m \, I_1(\beta,q)
   +
   \frac{4 \beta}{3} \, b m \Lambda \, I_3(\beta,q)
   \label{e:P-24a} \\
   & \qquad\qquad
   - 
   \frac{g M \beta}{3}
   - 
   \Bigl [\,
       b^2 + m^2 - \frac{1}{4} 
       - \frac{4}{9}\, b^2 m^2 \,
   \Bigr ] \, \frac{\beta}{2} \, I_2(\beta,q) \>,
   \notag \\
   v_{\theta}
   &=
   0 \>,
   \label{e:P-24b} \\
   v_q
   &=
   - 
   \frac{2 \beta}{3} \, M \, p b m \, f_1(\beta,q)
   -
   \frac{4 \beta}{3} \, M \, b m \Lambda \, f_2(\beta,q)
   \label{e:P-24c} \\
   & \qquad\qquad\qquad
   +
   M \,
   \Bigl [\,
       b^2 + m^2 - \frac{1}{4} 
       - \frac{4}{9}\, b^2 m^2 \,
   \Bigr ] \, \beta \, f_6(\beta,q) \>,
   \notag \\
   v_p
   &=
   2 M p 
   +
   \frac{2 M \beta}{3} \, b m \, I_1(\beta,q) \>,
   \label{e:P-24d} \\
   v_{\beta}
   &=
   \frac{2 M \beta}{3}
   -
   \frac{2 M \pi^2 \Lambda^2 }{3 \beta^3}
   +
   \frac{2 M}{3} \, p b m \, I_1(\beta,q)
   +
   \frac{4 M}{3} \, b m \Lambda \, I_3(\beta,q)
   \label{e:P-24e} \\
   & \qquad
   - 
   \frac{g M^2}{6}
   - 
   \frac{M}{2} \,
   \Bigl [\,
       b^2 + m^2 - \frac{1}{4} 
       - \frac{4}{9}\, b^2 m^2 \,
   \Bigr ] \,  I_2(\beta,q) 
   -
   \frac{4 M \beta}{3} \, p b m \, f_{10}(\beta,q)
   \notag \\
   & \qquad
   -
   \frac{8 M \beta}{3} \, b m \Lambda \, f_9(\beta,q)
   +
   \Bigl [\,
       b^2 + m^2 - \frac{1}{4} 
       - \frac{4}{9}\, b^2 m^2 \,
   \Bigr ] \, M \beta \, f_7(\beta,q) \>,
   \notag \\
   v_{\Lambda}
   &=
   \frac{2 \pi^2 M \Lambda}{3 \beta^2}
   +
   \frac{4 \beta M}{3} \, b m \, I_3(\beta,q) \>. 
   \label{e:P-24f}
\end{align}
\end{subequations}
The symplectic equations of motion are
\begin{equation}\label{e:P-29}
   \dot{Q}^{\mu}
   =
   f^{\mu\nu}(Q) \, u_{\nu}(Q) \>,
\end{equation}
from which we find:
\begin{subequations}\label{e:P-32}
\begin{align}
   \dot{M}
   &=
   - 2 M \, m b \beta f_1(\beta,q) \>,
   \label{e:P-32a} \\
   \dot{\theta}
   &=
   - 
   p^2
   +
   \frac{2}{3} \, \beta^2
   -
   \frac{5}{12} \, g \beta M
   +
   \frac{1}{3} \, m b p \beta \, I_1(\beta,q)
   +
   2 \, m b \beta \Lambda \, I_3(\beta,q)
   \label{e:P-32b} \\
   & \qquad
   +
   2 \, m b p \beta \, f_2(\beta,q)
   -
   \frac{2}{3} \, m b p \beta^2 \, f_{10}(\beta,q)
   -
   \frac{4}{3} \, m b \beta^2 \Lambda \,  f_9(\beta,q)
   \notag \\
   & \qquad
   -
   \frac{1}{4} \, 
   \Bigl [\,
       b^2 + m^2 - \frac{1}{4} 
       - \frac{4}{9}\, b^2 m^2 \,
   \Bigr ] \, \beta \, 
   [\, 3 \, I_2(\beta,q) - 2 \, \beta \, f_7(\beta,q) \,]  \, 
   \notag \\
   \dot{q}
   &=
   2 p 
   +
   \frac{2 \beta}{3} \, m b \, I_1(\beta,q)
   -
   2 \, m b \beta \, f_2(\beta,q) \>,
   \label{e:P-32c} \\
   \dot{p}
   &=
   \frac{2}{3} \, m b \beta \, p \, f_1(\beta,q)
   -
   \frac{8}{3} \, m b \beta \, \Lambda \, f_2(\beta,q)
   -
   \Bigl [\,
       b^2 + m^2 - \frac{1}{4} - \frac{4}{9}\, m^2 b^2 \,
   \Bigr ] \, \beta \, f_6(\beta,q) \, 
   \label{e:P-32d} \\
   \dot{\beta}
   &=
   - 
   m b \, \beta^2 f_1(\beta,q)
   -
   4 \beta \Lambda
   -
   \frac{8 \beta^4}{\pi^2} \, m b \, I_3(\beta,q)
   +
   \frac{12 \beta^4}{\pi^2} \, m b \, f_3(\beta,q) \>,
   \label{e:P-32e} \\
   \dot{\Lambda}
   &=
   -
   4 \Lambda^2
   +
   \frac{4 \beta^4}{\pi^2}
   +
   \frac{4}{\pi^2} \, \beta^3 p m b  \, I_1(\beta,q)
   +
   \frac{8}{\pi^2} \, \beta^3 \Lambda m b  \, I_3(\beta,q)
   \label{e:P-32f} \\
   & \qquad
   -
   \frac{g \beta^3 M}{\pi^2}
   -
   \frac{6 \beta^3}{\pi^2} \,
   \Bigl [\,
       b^2 + m^2 - \frac{1}{4} - \frac{4}{9}\, b^2 m^2 \,
   \Bigr ] \, f_8(\beta,q)
   \notag \\
   & \qquad
   -
   \frac{8 \beta^4}{\pi^2} \, b m \, p \, f_{10}(\beta,q)
   -
   \frac{16 \beta^4}{\pi^2} \, b m \Lambda \, f_9(\beta,q) \>.
   \notag
\end{align}
\end{subequations}
\end{widetext}
In Eq.~\ef{e:P-32f}, we use the identity \ef{e:ID-4}.
Here $M(t)$ is a dynamic variable. In order for the variational trial wave function to match the exact solution at $t=0$, the initial conditions are:
\begin{gather}\label{e:P-33}
   q_0 = 0
   \qc
   p_0 = 0
   \qc
   \beta_0 = 1
   \qc
   \Lambda_0
   =
   0
   \qc
   \theta_0 = - t \, 
   \\
   g M_0 
   = 
   \frac{(4 b^2 - 9)(4 m^2 - 9 )}{18} \>.
   \notag
\end{gather}
As a check, the right-hand-sides of Eqs.~\ef{e:P-32} vanish [except for $\dot{\theta}(0) = -1$] at these initial values, which guarantees that the exact solution is stationary.  For non-zero values of $q_0$ and/or $\beta_0$, the values of $p_0$ and $\Lambda_0$ are sometimes fixed by setting $\dot{q}_0 = 0$ and $\dot{\beta}_0=0$, and solving Eqs.~\ef{e:P-32c}, and \ef{e:P-32e} for $p_0$ and $\Lambda_0$, which gives:
\begin{subequations}\label{e:P-34}
\begin{align}
   p_0 
   &=
   \frac{1}{2} \,
   \Bigl [\,
      \dot{q}_0 
      - 
      \frac{2}{3} m b \beta_0 \, I_1(\beta_0,q_0) 
      \label{e:P-34a} \\
      & \qquad\qquad\qquad
      + 
      2 \, m b \beta_0 \, f_2(\beta_0,q_0) \,
   \Bigr ] \>,
   \notag \\
   \Lambda_0
   &=
   \frac{1}{4 \beta_0} \,
   \Bigl [\,
      -
      \dot{\beta}_0
      -
      m b \beta_0^2 f_1(\beta_0,q_0)
      \label{e:P-34b} \\
      & \qquad
      -
      \frac{8}{\pi^2} \, m b \beta^4 \, I_3(\beta_0,q_0)
      +
      \frac{12}{\pi^2} \, m b \beta^4 \, f_3(\beta_0,q_0) \,
   \Bigr ] \>.
   \notag
\end{align}
\end{subequations}

\begin{figure*}[t]
   \centering
   \subfigure[\ $m=0$]
   { \label{f:fig2a} \includegraphics[width=0.95\columnwidth]{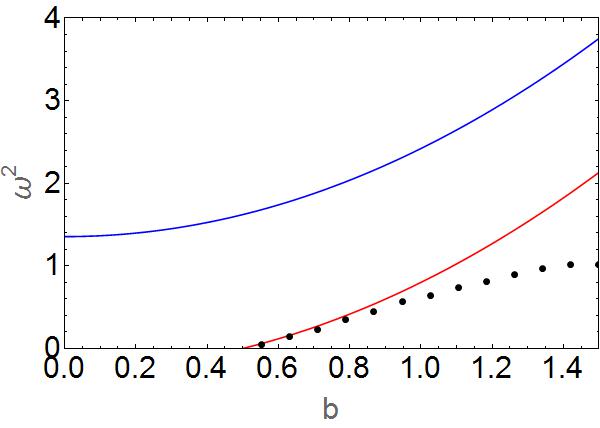} }
   \subfigure[\ $m=1.0$]
   { \label{f:fig2b} \includegraphics[width=0.95\columnwidth]{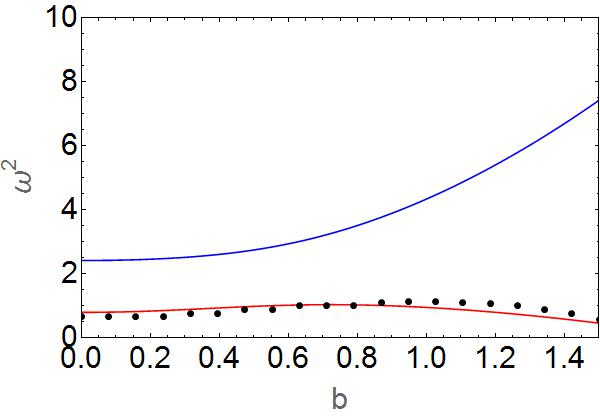} }
   \subfigure[\ $m=1.25$]
   { \label{f:fig2c} \includegraphics[width=0.95\columnwidth]{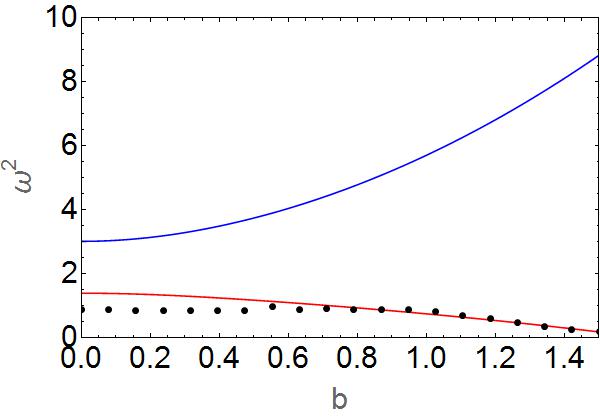} }
   \caption{\label{f:fig2}Plots of the linear response frequencies 
   $\omega_q^2$ (lower curve in red) and $\omega_{\beta}^2$ (upper
   curve in blue) as a function of $b$ for (a) $m=0$, (b) $m=1$, and (c) 
   $m=1.25$.
   The black dots represent data from the numerical simulation 
   (see Sec.~\ref{s:results}).  The product $mb$ controls the strength 
   of the imaginary part of the potential.}
\end{figure*}

When $m=0$, the external potential is real and $\dot M=0$.  The stability of the solutions to this equation for arbitrary $\kappa$ and for repulsive and attractive potential $V_1$ as well as positive and negative $g$ was studied using a variety of methods, and the stability properties and small oscillation frequencies for $q,p,\beta,\Lambda$ were determined in Ref.~\cite{r:Dawson:2017td}.  For that problem when we set $\kappa=1$ and $m=0$, our equations simplify to 
\begin{align}\label{e:reduce}
   \dot{q} 
   &= 2 \, p \>,
    \\
   \dot{\beta} 
   &= - 4 \beta \Lambda \>,
   \notag \\
   \dot{\Lambda } 
   &= - 4 \Lambda^2 + \frac{4 \beta^4}{\pi^2}  -
   \frac{g \beta^3 M}{\pi^2} 
   -
   \frac{6 \beta^3}{\pi^2} \,
   \Bigl [\,
       b^2  - \frac{1}{4} 
   \Bigr ] \, f_8(\beta,q)
   \notag 
\end{align}
which agrees with the results in Ref.~\cite{r:Dawson:2017td} once we use the fact that $f_3[G,q,\gamma]$ in that paper is just $\beta^2 f_8(\beta, q)$ here.  At $m=0$ the small oscillation equations for $\beta$ and $q$ decouple.  Using the expansions of the integrals found in Appendix A, we find that the small oscillation equations are:
\begin{align}\label{e:SOmzero}
   \delta \dot q &= 2 \delta p \, 
   \\
   \delta \dot p &= - \frac{8}{15} (b^2-1/4) \, \delta q \, 
   \notag
\end{align}
so that 
\begin{gather}\label{e:SOmzeroOmega}
 \delta \ddot q + \omega_q^2  \delta q = 0 \>,
 \\
 \omega_q^2 = \frac{16}{15} (b^2-1/4) \>.
 \notag
\end{gather}
This agrees with the result from the 4-parameter Ansatz.  However, we get a \emph{different} frequency for the $\beta$ oscillation,
\begin{align}\label{e:betaosc}
   \delta \dot \beta 
   &= 
   - 4 \delta \Lambda \>,
   \\
   \delta \dot \Lambda 
   &=
   \Bigl [\,
      \frac{4 b^2}{15} + \frac{4}{\pi^2} - \frac{1}{15} \,
   \Bigr ] \, \delta \beta \>,
   \notag
\end{align}
so that 
\begin{gather}\label{e:betaomega}
   \delta \ddot \beta + \omega_\beta^2 \, \delta \beta = 0 \>,
   \\
   \omega_\beta^2 
   =
   4 \, \Bigl [\,
      \frac{4 b^2}{15} + \frac{4}{\pi^2} - \frac{1}{15}\,
   \Bigr ] \>.
   \notag
\end{gather}
Plots of $\omega_q^2$ and $\omega_\beta^2$ for $m=0$ are shown in Fig.~\ref{f:fig2a}.

%
%
%
%

%
%
\section{\label{s:LinearRes}Linear response results for the six CC approximation}

We linearize the set of equations given in \ef{e:P-32} by expanding the equations about the exact solutions, $Q^{\mu} = Q_0^{\mu} + \delta Q^{\mu}$ keeping only the first order terms. Note that $Q_0^{\mu}$ are given in Eqs.~\ef{e:P-33}.  Using the expansions of Appendix~\ref{ss:expand}, we find
\begin{subequations}\label{e:E-2}
\begin{align}
   \delta \dot{M}
   &=
   - \frac{\pi}{2} \, m b \, M_0 \, \delta q \>,
   \label{e:E-2a} \\
   \delta \dot{\theta}
   &=
   -
   \frac{5}{12} \, g \delta M
   +
   \frac{7\pi}{18} \, m b \, \delta p
   \label{e:E-2b} \\
   & \quad
   +
   \frac{1}{3} \,
   \Bigl [\,
      1 
      + 
      \frac{2 \pi^2}{15}
      -
      \Bigl (\, \frac{1}{2} + \frac{\pi^2}{30} \, \Bigr ) \, g M_0 \,
   \Bigr ] \, \delta \beta \>,
   \notag \\
   \delta \dot{q}
   &=
   \frac{\pi}{72} \,
   \qty(\, 9 \pi^2 - 64 \,) m b \, \delta \beta + 2 \, \delta p \>,
   \label{e:E-2c} \\
   \delta \dot{p}
   &=
   \frac{4}{15} \, [\, g M_0 - 4 \,] \, \delta q
   -
   \frac{4 \pi}{9} \, m b \, \delta \Lambda \>,
   \label{e:E-2d} \\
   \delta \dot{\beta}
   &=
   \Bigl [\, \frac{\pi}{2} - \frac{20}{3 \pi} \, \Bigr ] \, 
   m b \, \delta q
   -
   4 \, \delta \Lambda \>,
   \label{e:E-2e} \\
   \delta \dot{\Lambda}
   &=
   \frac{2 b m}{3\pi} \delta p
   -
   \frac{ g }{\pi^2} \delta M 
   \label{e:E-2f} \\
   & \qquad
   +
   \frac{2}{15} 
   \Bigl [ \,
      - g M_0 + \frac{30}{\pi^2} + 4 \,
   \Bigr ] \, \delta \beta \>,
   \notag
\end{align}
\end{subequations}
where we have used the relation,
\begin{equation}\label{e:P-35}
   b^2 + m^2 - \frac{4}{9} \, b^2 m^2 - \frac{1}{4}
   =
   2 - \frac{1}{2} \, g M_0 \>.
\end{equation}

Equations~\ef{e:E-2} are written as
\begin{equation}\label{e:E-3}
   \delta \dot{Q}^{\mu}
   =
   M^{\mu}{}_{\nu}(Q_0) \, \delta Q^{\nu} \>,
\end{equation}
from which we find:
\begin{gather}\label{e:E-4}
   \delta \ddot{Q}^{\mu} 
   + 
   W^{\mu}{}_{\nu}(Q_0) \, \delta Q^{\nu}
   =
   0
   \\
   W^{\mu}{}_{\nu}(Q_0)
   =
   -
   M^{\mu}{}_{\sigma}(Q_0) M^{\sigma}{}_{\nu}(Q_0) \>.
   \notag
\end{gather}
Here $W^{\mu}{}_{\nu}(Q_0)$ is Hermitian.
The square of the linearized oscillation frequencies $\omega^2$ are given by the eigenvalues of $W^{\mu}{}_{\nu}(Q_0)$.  One can show that the matrix $W^{\mu}{}_{\nu}(Q_0)$ can be split into two blocks, one of them coupling $(\delta q, \delta \Lambda, \delta \theta)$, the other coupling $(\delta p,\delta \beta,\delta M)$.  Both of these blocks give identical eigenvalues, a zero eigenvalue and two non-zero eigenvalues.  For example, using Eqs.~\ef{e:E-2}, we find
\begin{subequations}\label{e:eigen-1}
\begin{align}
   \delta \ddot{q} - \qty[\, A \, \delta q + B \, \delta \Lambda \,] &= 0 \>,
   \label{e:eigen-1a} \\
   \delta \ddot{\Lambda} - \qty[\, D \, \delta q + E \, \delta \Lambda \,] &= 0 \>,
   \label{e:eigen-1b}
\end{align}
\end{subequations}
where
\begin{subequations}\label{e:eigen-2}
\begin{align}
   A
   &=
   \frac{8}{15} \, \qty( g M_0 - 4 ) 
   \label{e:eigen-2a} \\
   & \qquad\qquad
   + 
   \frac{(9 \pi^2 - 64)(3 \pi^2 - 40) \, b^2 m^2}{432} \>,
   \notag \\
   B
   &=
   \qty( 16 - 3 \pi^2 ) \, \frac{\pi \, b m}{6} \>,
   \label{e:eigen-2b} \\
   D
   &=
   b m \,
   \Bigl \{ \,
      \frac{g M_0}{2 \pi}
      +
      \frac{2( 3 \pi^2 - 40 )}{3 \pi^3}
      \label{e:eigen-2c} \\
      & \qquad\qquad
      +
      \frac{(g M_0 - 4)(16 - \pi^2 )}{15 \pi} \,
   \Bigr \} \>,
   \notag \\
   E
   &=
   -
   \frac{16}{\pi^2} - \frac{8}{27} \, m^2 b^2 + \frac{8}{15}\, ( g M_0 - 4 ) \>,
   \label{e:eigen-2d}
\end{align}
\end{subequations}
from which we find
\begin{equation}\label{e:eigen-3}
   \omega^2
   =
   \frac{1}{2} \,
   \qty[\,
      - \qty(A + E) \pm \sqrt{ \qty(A - E)^2 + 4 \, B D } \, ] \>.
\end{equation}

Although these two frequencies increase together when  $m=0$ as a function of $b$, once we get near  $m=1$ they start repelling each other and the dependence of the lower frequency has a maximum as a function of $b$ instead of monotonically increasing.  This is shown in Fig.~\ref{f:fig2}.  Note that when $m=0$ and $b^2< 1/4$, the potential becomes repulsive, which leads to $\omega^2 <0$ and thus to a translational instability.  This was studied in detail in Ref.~\cite{r:Dawson:2017td}.

%

%
%
\section{\label{s:Numerical}Computational Strategy}

In our previous sections we were able to develop a six parameter variational approach to the time evolution of slightly perturbed solutions of the NLS equation in an external complex potential.  We were able to get an explicit analytic  expression as a function of $m, b$, of two oscillation frequencies that affect the response of the solution to  small perturbations.  So the first question we would like to answer is how does this analytic result compare to the actual response found by numerically solving the NLS equation. The second question we want to answer is the domain of applicability of the variational approach in terms of predicting the actual time evolution of the low order moments of the solution.  This has two parts: (i)  for fixed $m,b$ how long is the approximation valid and (ii) as we increase the size of the complex part of the potential, by say varying $m$ for fixed $b$, when does this approach start losing its validity. In our approximation for all $b,m$ that correspond to an attractive potential, there is no translational instability.  So we would like to see in our numerical simulations, that for the case $m=1$ (and $\kappa=1$), the translational instability that arises due to mixing of the solution we are considering with the first excited state in the potential occurs at times much later than the domain of applicability of the six CC method.  For that case when $0.56 < b < 1.37$ a late time translational instability was found. 

To study numerically the evolution of Eqs.~\ef{eqn1}, we have used a homemade code using a Crank-Nicolson scheme \cite{ref:numrec}. In Ref.~\cite{PhysRevE.94.032213} we have shown that the Crank-Nicolson scheme is a reliable method for successfully solving Eq.~\ef{eqn1} in the presence of a complex potential. For the sake of comparison with the analytical calculations, the initial soliton shape $\psi(x,0)$ in our simulations is given by Eqs.~\ef{e:T-1} and \ef{e:T-2} at $t=0$.  
The complex soliton shape in the transverse spatial domain $x$ was represented in a regular grid with mesh size $\Delta x=2\times 10^{-6}$ and free boundary conditions were imposed. The mesh size was chosen to be much smaller than the initial soliton width parameter $1/\beta(0)=1$, so that very small variations of the soliton position could be accurately measured by using a center of mass definition, \ie\ $q=\expval{x}$, where the expectation value is defined in Eq.~\ef{e:evdef}. 
The soliton width $W(t)$ is the square root of the normalized second moment $G_2 = \pi^2/(12 \beta^2(t))$.  The soliton width parameter $1/\beta(t)$ in the simulations was calculated by using the expression $1/\beta(t) = \sqrt{G_2(t)/G_2(0)}$.  The other CC's measured in the simulations were the amplitude $A(t)=\max_{x \in \R} \sqrt{\rho(x,t)}$ and the mass $M(t)$ given by Eq.~\ef{eqMass1}.


%
%
\section{\label{s:results}Comparison of collective variable theories with simulations}

Our potential is symmetric in $b \leftrightarrow m$.   When either $b$ or $m=0$ the potential is real and the small oscillation equations for $q, p$ $\beta, \Lambda$ decouple giving rise to separate oscillation frequencies in that regime. Once the imaginary part turns on, we expect that these two oscillation frequencies appear to a certain degree in all the collective coordinates.  Note  that in the collective coordinate approach the mass is  related to the height and the CC parameter $\beta$ and is not an independent parameter, namely  $M(t) = 2 A^2(t)/\beta(t)$.  First let us choose $g=1, \kappa=1, m=0$ and $b=1 $ to see how well our CC approximation works when compared with numerical simulations when the potential is real. For our simulations we choose the parameters $g = 1, \kappa = 1, q_0=0.001, \beta = 1.001$ and all other parameters those of the exact solution.  The small oscillation theory for this case predicts separate oscillation frequencies for $q$ and $\beta$, namely $T_q= 7.025$, $\omega_q^2 = 0.800$ and $T_\beta=4.038$, $\omega_{\beta}^2=2.421$.  These frequencies are located on the two branches in Fig.~\ref{f:fig2a}, and agree with the six CC approximation.  The simulation results are represented by the black data points.  This is seen in both the six CC approximation and the numerical simulation.  

\begin{figure}[t]
   \centering
  \subfigure[\ Position $q(t)$ \vs\ $t$]
  { \label{f:fig3a} 
  \includegraphics[width=0.95\columnwidth]{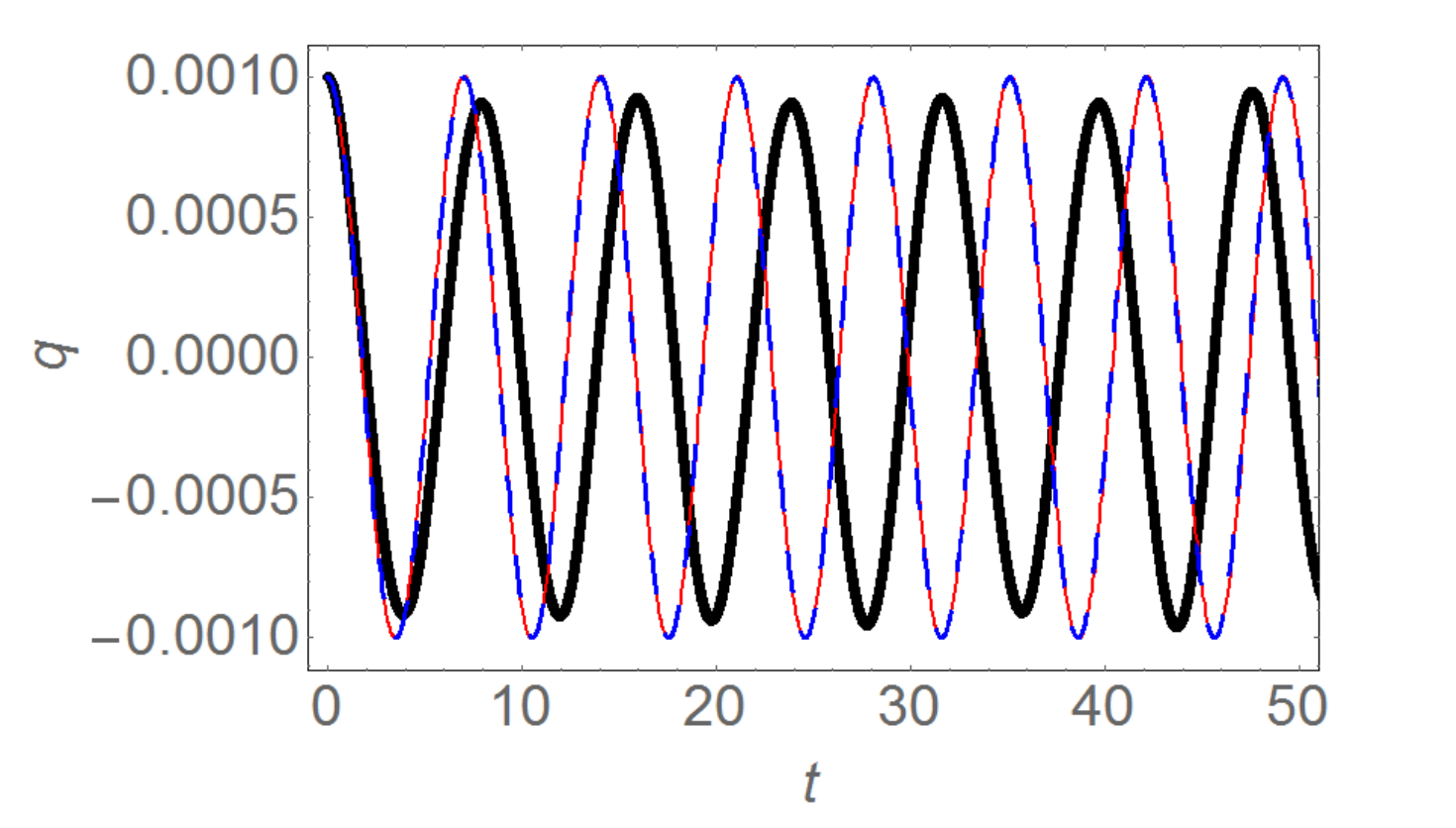} }
  \subfigure[\ Width $1/ \beta(t)-1/\beta(0)$ \vs $t$]
  { \label{f:fig3b} 
  \includegraphics[width=0.95\columnwidth]{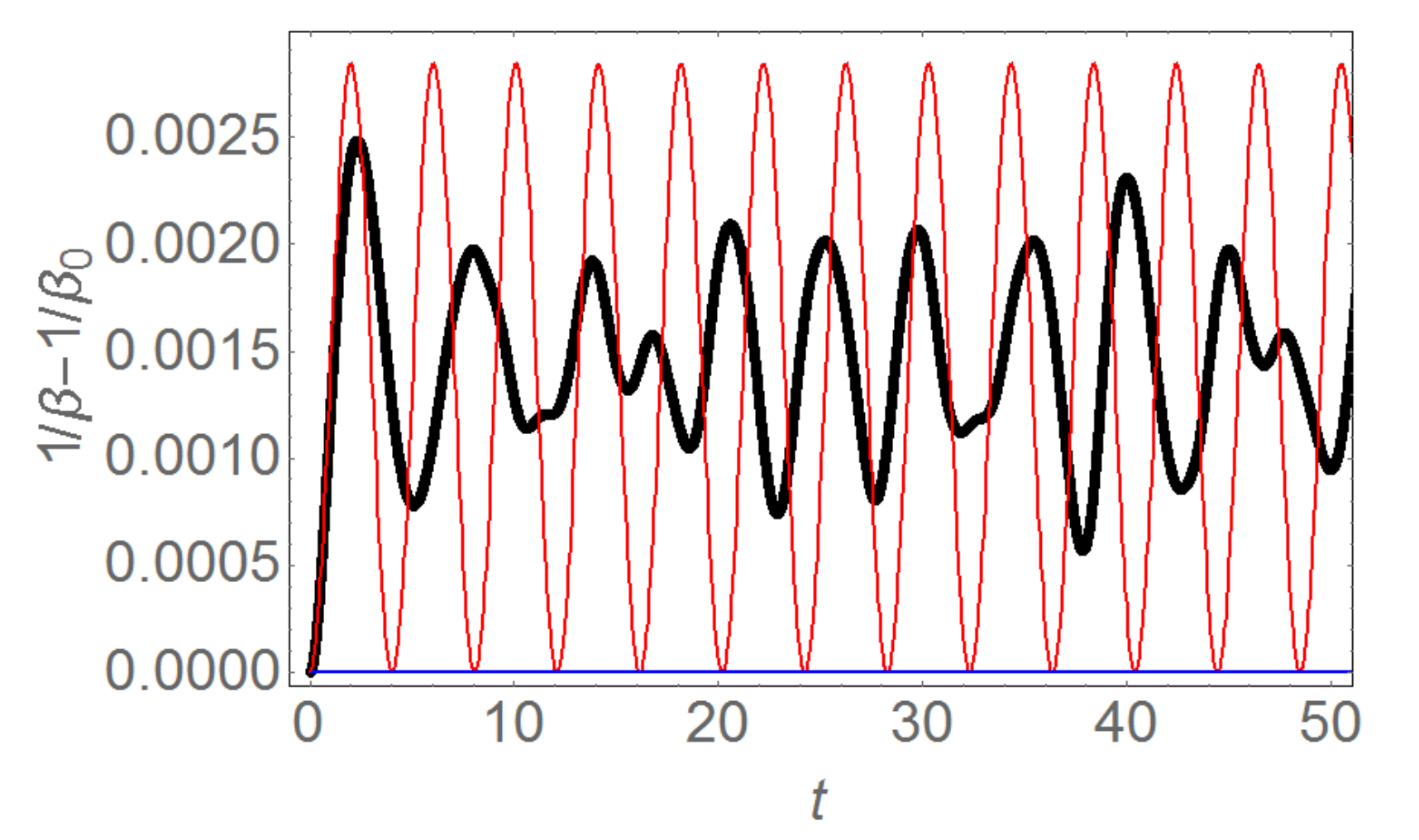} }
  \caption{\label{f:fig3} Comparison of the four CC (blue line), 
  six CC (red line) and numerical simulation (black line). 
  Parameters and initial conditions are $m=0, q_0=0.001, \beta =1.001$.
  All other initial conditions are the exact solution values. 
  Since $M=M_0$, we display only $q$ and  $1/\beta-1/\beta_0$.  
  For this choice the two linear response periods are $T_q = 7.025$, 
  and $T_\beta=4.038$.}
\end{figure}

Since the perturbation is so small, we subtract the initial value of $1/\beta_0=1$ from $1/\beta$ to show the oscillation in the numerical simulations.  We see that for $q(t)$, both the amplitude as well as period of oscillation are well reproduced by the six CC theory.  This is shown in Fig.~\ref{f:fig3}.  

For the width parameter $1/\beta$, the oscillation period is $4.00$ which agrees well with the linear response result $4.038$, but not so well with the simulation result.  Here the spectrum consists of several peaks around the frequency of $1.132$, which corresponds to the period $5.55$.  Moreover, the soliton amplitude $A(t)$ has the period $3.85$, which is rather close to the above value of $4.00$.

%
%
%
%

For our simulations with a complex potential we choose the parameters $g = 1, \kappa = 1, m = 1$, and three values for $b$. First we choose $ b = 0.1$ so  that the imaginary part of the potential is small, and the dissipation is weak. Next we choose  $b = 0.5$ which is located in the lower stability regime $0 < b < 0.56$; and $b = 1.45$ is located in the upper stability regime $1.37 < b < 1.5$.

The exact solution Eq.~\ef{psi0wf} is stationary and is obtained by the CC Ansatz Eq.~\ef{e:T-1} with the initial conditions (ICs) $q_0 = 0, p_0 = 0, \beta_0 = 1, \Lambda_0 = 0, \theta_0 = 0$,  and $g M_0 = (4 b^2-9)(4 m^2-9)/18$, see Eqs.~\ef{e:P-33}. In order to test the stability of the exact solution, we choose ICs that are slightly different from the above values.  This excites intrinsic oscillations of the soliton which are seen in the time evolution of the CCs, which is obtained by solving the six CC equations, Eqs.~\ef{e:P-32}, by a Mathematica program. These oscillations are compared with the oscillations which are observed in the simulations, i.e. in the numerical solution of the NLS equation. In particular, the frequencies, periods, and amplitudes of the oscillations are compared.

\begin{figure}[t]
   \centering
   \subfigure[\ Position $q(t)$ \vs\ $t$]
   { \label{f:fig4a} 
   \includegraphics[width=0.95\columnwidth]{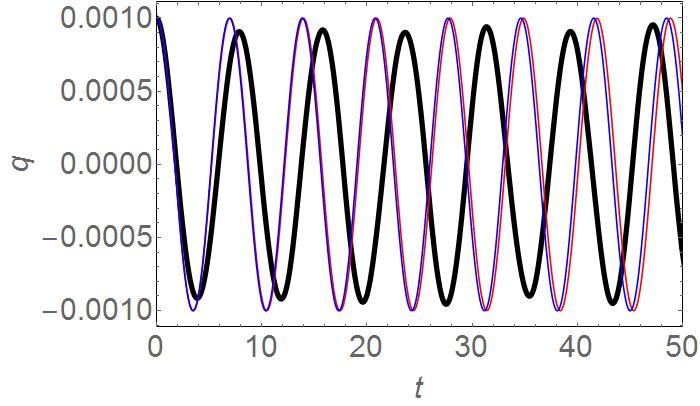} }
   \subfigure[\ Amplitde $A(t)$ \vs\ $t$]
   { \label{f:fig4b} 
   \includegraphics[width=0.95\columnwidth]{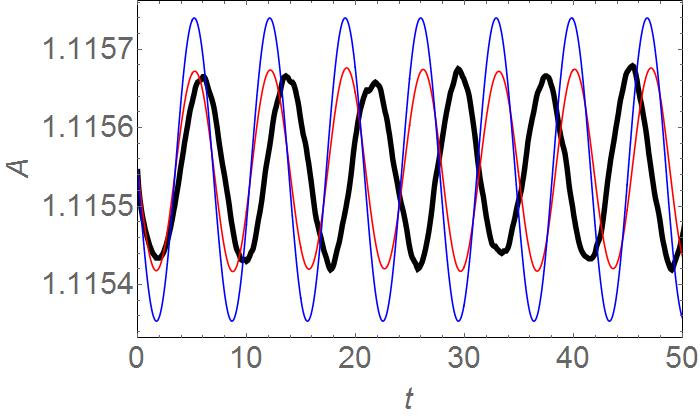} } 
   \subfigure[\ Width $1/ \beta(t)$ \vs\ $t$]
   { \label{f:fig4c} 
   \includegraphics[width=0.95\columnwidth] {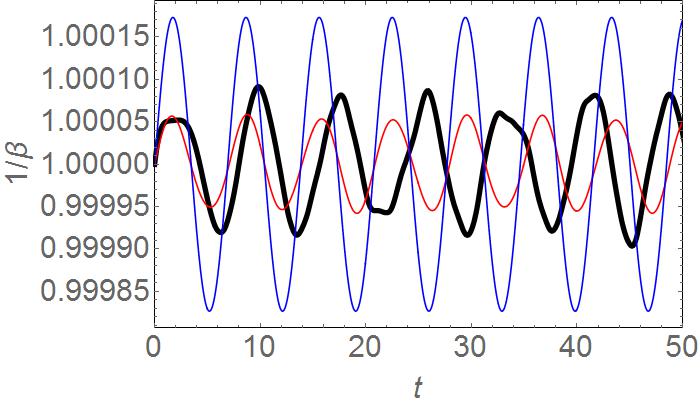} }
   \caption{\label{f:fig4}Comparison of the four CC (blue line), 
   six CC (red line), and numerical simulation (black line) for $b=0.1$.
   Parameters and initial conditions are $m=1, q_0=.001$,
   all other initial conditions are the exact solution values.  
   Here we display $q(t)$, $A(t)$, and $1/\beta(t)$.  
   For this choice the two linear response periods are: $T_q = 7.025$ 
   and $T_\beta=  4.038$.}
\end{figure}

For the case $b = 0.1$ the four CC and six CC results are nearly identical and agree very well with the simulation results in Fig.~\ref{f:fig4}. 
%
%
%
%
The periods of the oscillations are $T_{4CC} = T_{6CC} = 7.14$, compared to $T_{\text{sim}} = 7.69$. This means that the error in the CC theories is only 7\%.

\begin{figure}[t]
   \centering
  \subfigure[\ $q(t)$ \vs\ $t$]
  { \label{f:fig5a} 
  \includegraphics[width=0.95\columnwidth]{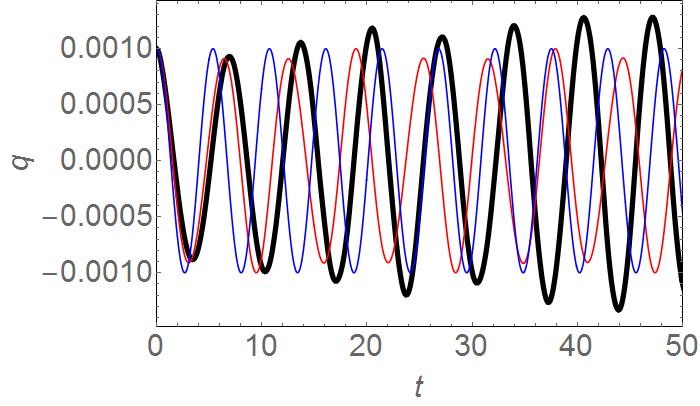} }
  \subfigure[\ $ A(t) $]
   { \label{f:fig5b} 
   \includegraphics[width=0.95\columnwidth]{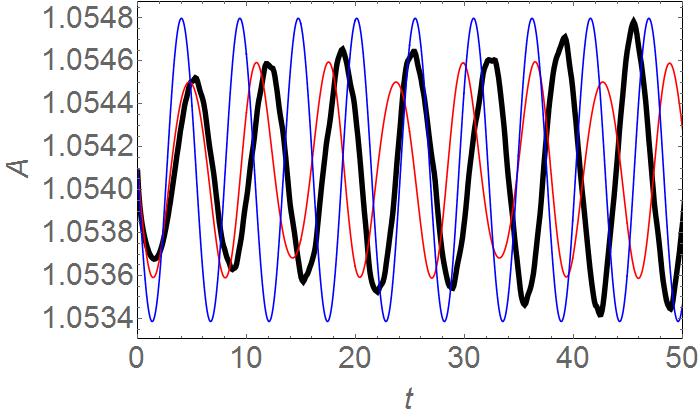} } 
  \subfigure[\ $1/ \beta(t)$]
  { \label{f:fig5c} 
  \includegraphics[width=0.95\columnwidth] {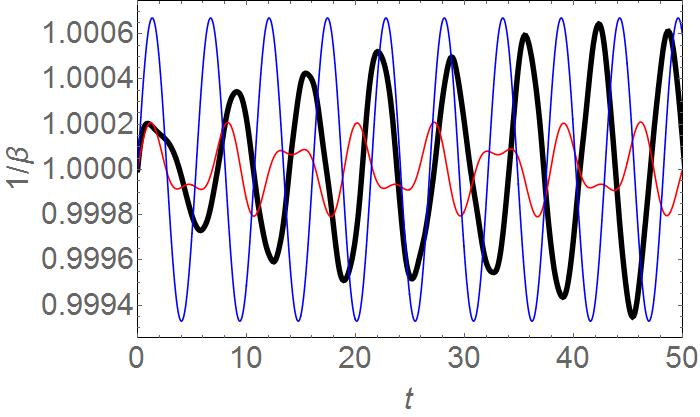} }
   \caption{\label{f:fig5}Comparison of the four CC (blue line), 
   six CC (red line), and numerical simulation (black line) for $b=0.5$. 
   Other parameters and initial conditions being the same as in Fig. 4.}
\end{figure}

For the case $b = 0.5$ the six CC result is much better than the four CC result and agrees rather well with the simulation shown in Fig.~\ref{f:fig5}.  The periods are $T_{4CC} = 5.26$, $T_{6CC} = 6.25$ and $T_{\text{sim}} = 6.67$, the error is 6\%.
%
%
%
%

\begin{figure*}[t]
  \centering
  \subfigure[\ $q(t)$ \vs\ $t$]
  { \label{f:fig6a} 
  \includegraphics[width=0.95\columnwidth]{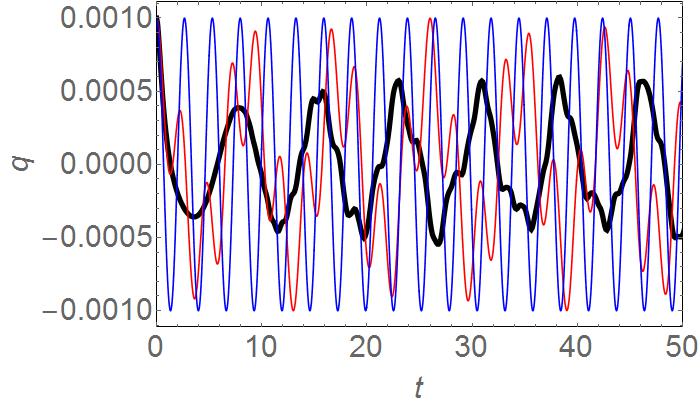} }
  \subfigure[\ $ A(t) $]
  { \label{f:fig6b} 
  \includegraphics[width=0.95\columnwidth]{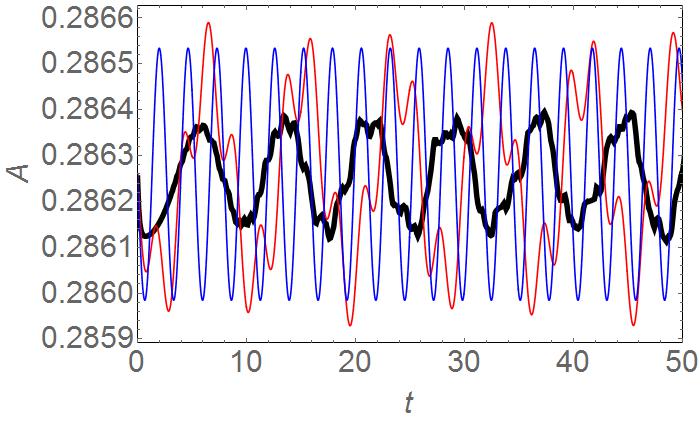} } 
  \subfigure[\ $1/ \beta(t)$]
  { \label{f:fig6c} 
  \includegraphics[width=0.95\columnwidth]{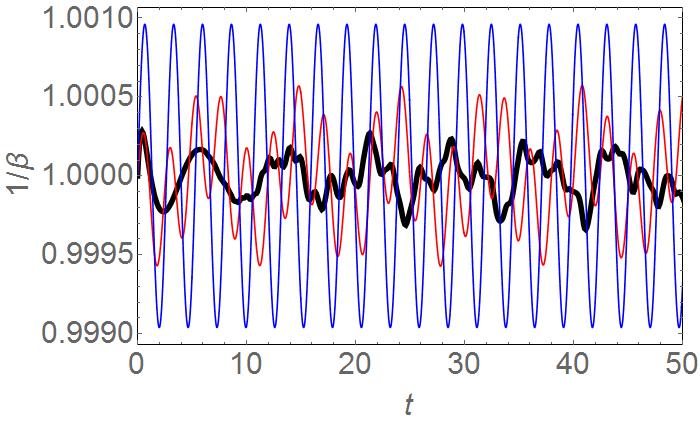} }
  \caption{\label{f:fig6}Comparison of the four CC (blue line), 
   six CC (red line), and numerical simulation (black line) for $b=1.45$. 
   Other parameters and initial conditions being the same as in Fig. 4.}
\end{figure*}

For the case $b = 1.45$ the four CC result poorly fits the numerical result.  The six CC result is very anharmonic and the oscillation amplitudes do not agree well with the simulations  as seen in Fig.~\ref{f:fig6}.  Nevertheless, the periods $T_{6CC} = 8.33$ and $T _{\text{sim}} = 7.69$ agree within an error of 8\%.  Interestingly, the spectra exhibit a second frequency which is obtained also in the linear response theory.  Fig.~\ref{f:fig2b} shows the two frequencies for all values of $b$. However, the simulations show only one frequency.  
%
%

So far we have always taken $q_0 = 0.001$, and the other ICs as in the exact solution. Choosing a finite value for $p_0$ gives very similar results, because the $q$ and $p$ oscillations are related, see the relations below Eq.~\ef{e:SOmzero}.  Let us now consider $b=1.45$ and finite values for $\Lambda_0$ which will also affect the width $1/\beta$ because their oscillations are related. Choosing a very small, negative value $\Lambda_0 = -0.00005$, and increasing this value by steps, we find that the anharmonicity of the CCs gradually decreases. For $\Lambda_0 = -0.00025$ the oscillations are nearly harmonic and the periods are the same as in Fig.~\ref{f:fig6}

For $\Lambda_0 = +0.00025$ the periods are again the same as in Fig.~\ref{f:fig6}.  However, the spectrum of $M(t)$ exhibits a second peak at $T_2 = 2.38$ which is stronger than the first peak at $T_1 = 8.33$.  This second peak belongs to the upper branch in Fig.~\ref{f:fig2b} which was obtained by our linear response theory.  However, this peak is not seen in the simulations.

%
%
\section{\label{s:conclusions}Conclusions}

In this paper we investigated the domain of applicability of a four and six collective coordinate approximation to study the response of  the nodeless solution of the NLS equation in the presence of a complex potential to small perturbations.  This type of approximation had been used in the past to study the response of exact solutions of the NLS equation when in the presence of weak harmonic complex external potentials. In this paper we instead considered a $\PT$-symmetric potential where we could vary the strength of the complex part of the potential from zero to its maximum allowed value.  Using a small oscillation approximation to the CC equations we were able to obtain analytic expressions for the two frequencies of small oscillation found in our six CC approximation.  These frequencies were quite close to those that were found in the numerical simulations of the discretized  PDEs when we perturbed the initial conditions of the exact solution. This was true for all allowed values of the parameter product  $bm$ which governed the strength of the imaginary part of the potential.  We found that as we increased $bm$, the four CC approximation quickly broke down.  The six CC approximation was quite a reasonable approximation even at $bm=1/2$, but at the maximum value we studied $bm=1.45$, it tracked accurately the position of the
solitary wave for less than 1/4 of a period and then began to differ from the numerical solution. 

%
%
\acknowledgments

F.C.\ would like to thank the Santa Fe Institute  and the Center for Nonlinear Studies at Los Alamos National Laboratory for their hospitality. 
F.G.M. \ and N.R.Q.\ acknowledge financial support from the Ministerio de Econom\'{\i}a y Competitividad (Spain) through FIS2014-54497-P.  
F.G.M.\ also acknowledges financial support from the Plan Propio of Universidad de Seville and is grateful for the hospitality of the Mathematical Institute of the University of Seville (IMUS) and of the
Theoretical Division and Center for Nonlinear Studies at
Los Alamos National Laboratory.
N.R.Q.\ also acknowledges financial support from the Junta de Andalucia (Spain) under Projects No.~FQM207 and the Excellent Grant P11-FQM-7276.
E.A.\ gratefully acknowledges support from the Fondo Nacional
de Desarrollo Cientifico y tecnologico (FONDECYT) project No. 1141223
and from the Programa Iniciativa Cientfica Milenio (ICM)  Grant No. 130001. 
A.K.\ is grateful to Indian National Science Academy (INSA) for awarding him INSA Senior Scientist position at Savitribai Phule Pune
University, Pune, India. 
B.M.\ and J.F.D.\ would like to thank the Santa Fe Institute for their hospitality.  
B.M.\ acknowledges support from the National Science Foundation through its employee IR/D program.
The work of A.S.\ was supported by the U.S.\ Department of Energy. 

%
%
\appendix

%
%
\section{\label{s:integrals}Definition of integrals}

\begin{widetext}
We note that
\begin{subequations}\label{e:I-1}
\begin{align}
   \dv{}{z} \sech^2(z)
   &=
   - 2 \sech^2(z) \, \tanh(z)
   =
   - 2 \sech^3(z) \, \sinh(z) \>,
   \label{e:I-1a} \\
   \dv{}{z} \tanh(z)
   &=
   \sech^2(z) \>.
   \label{e:I-1b}
\end{align}
\end{subequations}
Some useful integrals are the following:
\begin{subequations}\label{e:I-2}
\begin{align}
   \tint \dd{z} \sech^2(z)
   &=
   2 \>, 
   \label{e:I-2a} \\
   \tint \dd{z} \sech^3(z)
   &=
   \frac{\pi}{2} \>, 
   \label{e:I-2b} \\
   \tint \dd{z} \sech^4(z)
   &=
   \frac{4}{3} \>, 
   \label{e:I-2c} \\
   \tint \dd{z} z^2 \sech^2(z)
   &=
   \frac{\pi^2}{6} \>, 
   \label{e:I-2d} \\
   \tint \dd{z} \sech^2(z) \tanh^2(z)
   &=
   \frac{2}{3} \>.
   \label{e:I-2e}
\end{align}
\end{subequations}
We define:
\begin{subequations}\label{e:I-3}
\begin{align}
   I_1(\beta,q)
   &=
   \tintx \sech^2( \beta y ) \sech(x)
   =
   \tinty \sech^2( \beta y ) \sech(y+q) \>, 
   \label{e:I-3a} \\
   I_2(\beta,q)
   &=
   \tintx \sech^2( \beta y ) \sech^2(x)
   =
   \tinty \sech^2( \beta y ) \sech^2(y+q) \>, 
   \label{e:I-3b} \\
   I_3(\beta,q)
   &=
   \tintx y \sech^2( \beta y ) \sech(x)
   =
   \tinty y \sech^2( \beta y ) \sech(y+q) \>. 
   \label{e:I-3c}
\end{align}
\end{subequations}
Also, we define:
\begin{subequations}\label{e:I-4}
\begin{align}
   f_1(\beta,q)
   &=
   \tint \dd{y} \sech^2(\beta y) \, \sech(y + q)  \, \tanh(y + q) \>,
   \label{e:I-4a} \\
   f_2(\beta,q)
   &=
   \tint \dd{y} y \, \sech^2(\beta y) \, \sech(y + q) \, \tanh(y + q) \>,
   \label{e:I-4b} \\
   f_3(\beta,q)
   &=
   \tint \dd{y} y^2 \, \sech^2(\beta y) \, \sech(y + q) \, \tanh(y + q) \>,
   \label{e:I-4c} \\
   f_4(\beta,q)
   &=
   \tint \dd{y} \sech^3(\beta y) \, \sech(y + q) \, \tanh(y + q) \>,
   \label{e:I-4d} \\
   f_5(\beta,q)
   &=
   \tint \dd{y} y \, \sech^3(\beta y) \, \sech(y + q) \, \tanh(y + q) \>,
   \label{e:I-4e} \\
   f_6(\beta,q)
   &=
   \tint \dd{y}
   \sech^2(\beta y) \,
   \sech^2(y + q) \, \tanh(y + q) \>,
   \label{e:I-4f} \\
   f_7(\beta,q)
   &=
   \tint \dd{y} y \,
   \sech^2(\beta y) \, \tanh(\beta y) \,
   \sech^2(y + q) \>,
   \label{e:I-4g} \\
   f_8(\beta,q)
   &=
   \tint \dd{y} y \,
   \sech^2(\beta y) \, \sech^2(y + q) \, \tanh(y+q) \>, 
   \label{e:I-4h} \\
   f_9(\beta,q)
   &=
   \tint \dd{y} y^2 \,
   \sech^2(\beta y) \, \tanh(\beta y) \, \sech(y + q) \>, 
   \label{e:I-4i} \\
   f_{10}(\beta,q)
   &=
   \tint \dd{y} y \,
   \sech^2(\beta y) \, \tanh(\beta y) \, \sech(y + q) \>.
   \label{e:I-4j}
\end{align}
\end{subequations}
Partial derivatives of $I_1(\beta,q)$ are given by 
\begin{subequations}\label{e:I-5}
\begin{align}
   \pdv{I_1(\beta,q)}{q}
   &=
   - \tint \dd{y}
   \sech^2(\beta y) \, \sech(y + q) \, \tanh(y + q)
   =
   - f_1(\beta,q) \>,
   \label{e:I-5a} \\
   \pdv{I_1(\beta,q)}{\beta}
   &=
   - 2 \tint \dd{y} y \,
   \sech^2(\beta y) \, \tanh(\beta y) \, \sech(y + q)
   =
   - 2 f_{10}(\beta,q) \>.
   \label{e:I-5b} 
\end{align}
\end{subequations}
Partial derivatives of $I_2(\beta,q)$ are given by 
\begin{subequations}\label{e:I-6}
\begin{align}
   \pdv{I_2(\beta,q)}{q}
   &=
   - 2 \tint \dd{y}
   \sech^2(\beta y) \, \sech^2(y + q) \, \tanh(y + q)
   =
   - 2 f_6(\beta,q) \>,
   \label{e:I-6a} \\
   \pdv{I_2(\beta,q)}{\beta}
   &=
   - 2 \tint \dd{y} y \,
   \sech^2(\beta y) \, \tanh(\beta y) \, \sech^2(y + q)
   =
   - 2 f_7(\beta,q) \>.
   \label{e:I-6b} 
\end{align}
\end{subequations}
Partial derivatives of $I_3(\beta,q)$ are given by 
\begin{subequations}\label{e:I-7}
\begin{align}
   \pdv{I_3(\beta,q)}{q}
   &=
   - \tint \dd{y} y \,
   \sech^2(\beta y) \, \sech(y + q) \, \tanh(y + q)
   =
   - f_2(\beta,q) \>,
   \label{e:I-7a} \\
   \pdv{I_3(\beta,q)}{\beta}
   &=
   - 2 \tint \dd{y} y^2 \,
   \sech^2(\beta y) \, \tanh(\beta y) \, \sech(y + q)
   =
   - 2 f_9(\beta,q) \>.
   \label{e:I-7b} 
\end{align}
\end{subequations}
A useful identity is obtained by integration of $f_7(\beta,q)$ by parts.  Using
\begin{equation}\label{e:ID-2}
   \pdv{y} \, \sech^2(\beta y)
   =
   - 2 \beta \, \sech^2(\beta y) \tanh(\beta y) \>,
\end{equation}
we find
\begin{align}\label{e:ID-3}
   - 2 \beta f_7(\beta,q)
   &=
   \tint
   y \, \sech^2(y + q) \,
   \dd{ \qty{ \sech^2(\beta y) } }
   \\
   &=
   -
   \tint
   \sech^2(\beta y) \, \dd{ \qty{ y \, \sech^2(y + q) \, } } 
   \notag \\
   &=
   -
   \tint \dd{y} \sech^2(\beta y) \sech^2(y + q)
   +
   2 \tint \dd{y} \sech^ 2(\beta y) \sech^2(y + q) \tanh(y + q) 
   \notag \\
   &=
   - I_2(\beta,q) + 2 \, f_8(\beta,q) \>.
   \notag
\end{align}
That is, 
\begin{equation}\label{e:ID-4}
   I_2(\beta,q) - 2 \beta f_7(\beta,q)
   =
   2 \, f_8(\beta,q) \>.
\end{equation}
We use this identity in the $\dot{\Lambda}$ equation, \ef{e:P-32f}.  Next, we now consider the expansion of the integrals and find to first order:
%
%
\begin{subequations}\label{e:E-1}
\begin{align}
   I_1(1+\delta\beta,\delta q)
   &=
   \frac{\pi}{2} - \frac{\pi}{3} \, \delta \beta \>,
   \label{e:E-1a} \\
   I_2(1+\delta\beta,\delta q)
   &=
   \frac{4}{3} - \frac{2}{3} \, \delta \beta \>,
   \label{e:E-1b} \\
   I_3(1+\delta\beta,\delta q)
   &=
   - \frac{\pi}{6} \, \delta q \>,
   \label{e:E-1c} \\
   f_1(1+\delta\beta,\delta q)
   &=
   \frac{\pi}{4} \, \delta q \>,
   \label{e:E-1d} \\
   f_2(1+\delta\beta,\delta q)
   &=
   \frac{\pi}{6} + \frac{\pi}{48} (16 - 3 \pi^2) \, \delta \beta \>,
   \label{e:E-1e} \\
   f_3(1+\delta\beta,\delta q)
   &=
   \frac{\pi}{48} ( -32 + 3 \pi^2 ) \, \delta q \>,
   \label{e:E-1f} \\
   f_6(1+\delta\beta,\delta q)
   &=
   \frac{8}{15} \, \delta q \>,
   \label{e:E-1g} \\
   f_7(1+\delta\beta,\delta q)
   &=
   \frac{1}{3} + \frac{2}{45} ( -15 + \pi^2 ) \, \delta \beta \>,
   \label{e:E-1h} \\
   f_8(1+\delta\beta,\delta q)
   &=
   \frac{1}{3} - \frac{2 \pi^2}{45} \, \delta \beta \>,
   \label{e:E-1i} \\
   f_9(1+\delta\beta,\delta q)
   &=
   \frac{\pi}{96} ( 16 - 3 \pi^2 ) \, \delta q \>,
   \label{e:E-1j} \\
   f_{10}(1+\delta\beta,\delta q)
   &=
   \frac{\pi}{6} + \frac{\pi}{32} ( - 16 + \pi^2 ) \, \delta \beta \>. 
   \label{e:E-1k}
\end{align}
\end{subequations}
\end{widetext}

%
%
\section{\label{s:TW}Generalized traveling wave method}

This method was named and used in a paper by Quintero, Mertens and Bishop \cite{PhysRevE.82.016606}.  We will show here that it is an alternative way to obtain Eq.~\ef{e:P-29} for the rate of change of the collective coordinates.  The authors substitute the trial wave function directly into \Schrodinger's equation.  This gives
\begin{subequations}\label{e:TW-1}
\begin{align}
   &\rmi \, \dot{Q}^{\nu} \partial_{\nu} \tpsi^{\phantom\ast}(x,Q)
   +
   \tpsi_{xx}^{\phantom\ast}(x,Q)
   +
   g \, |\tpsi(x,Q)|^{2\kappa} \, \tpsi^{\phantom\ast}(x,Q)
   \notag \\
   &=
   [\, V_1(x) + \rmi V_2(x) \, ] \, \tpsi^{\phantom\ast}(x,Q) \>,
   \label{e:TW-1a} \\
   - &\rmi \, \dot{Q}^{\nu} \partial_{\nu} \tpsi^{\ast}(x,Q)
   +
   \tpsi_{xx}^{\phantom\ast}(x,Q)
   +
   g \, |\tpsi(x,Q)|^{2\kappa} \, \tpsi^{\ast}(x,Q)
   \notag \\
   &=
   [\, V_1(x) - \rmi V_2(x) \, ] \, \tpsi^{\ast}(x,Q) \>.
   \label{e:TW-1b}   
\end{align}
\end{subequations}
Multiply \ef{e:TW-1a} by $\partial_{\mu} \tpsi^{\ast}(x,Q)$ and \ef{e:TW-1b} by $\partial_{\mu} \tpsi(x,Q)$ and add them to give
\begin{align}\label{e:TW-2}
   &\rmi \,
   \{\, 
      [\partial_{\mu} \tpsi^{\ast}] \, 
      [\partial_{\nu} \tpsi]
      -
      [\partial_{\nu} \tpsi^{\ast}] \, 
      [\partial_{\mu} \tpsi] \,
   \} \, \dot{Q}^{\nu}
   +
   [ \partial_{\mu} \tpsi^{\ast}] \,
   \tpsi_{xx}
   \\
   & \!\!\!\!
   +
   [\partial_{\mu} \tpsi] \,
   \tpsi_{xx}^{\ast}
   +
   \{\,
      g \, |\tpsi|^{2\kappa}
      -
      V_1(x) \,
   \} \,
   \{\, 
      [\partial_{\mu} \tpsi^{\ast}] \, \tpsi
      +
      [\partial_{\mu} \tpsi] \, \tpsi^{\ast} \,
   \}
   \notag \\
   & \quad
   =
   \rmi \, V_2(x) \,
   \{\, 
      [\partial_{\mu} \tpsi^{\ast}] \, \tpsi
      -
      [\partial_{\mu} \tpsi] \, \tpsi^{\ast} \,
   \} \>.   
   \notag
\end{align}
Integrating \ef{e:TW-2} over $x$ and the second term by parts gives
\begin{equation}\label{e:TW-3}
   I_{\mu\nu}(Q) \, \dot{Q}^\nu
   =
   \partial_\mu H(Q) + R_{\mu}(Q) \>,
\end{equation}
where
\begin{subequations}\label{e:TW-4}
\begin{align}
   I_{\mu\nu}(Q)
   &=
   \rmi
   \!\tint \dd{x} \!
   \bigl \{
      [\, \partial_\mu \tpsi^{\ast} \,]\,
      [\, \partial_\nu \tpsi \,]
      -
      [\, \partial_\nu \tpsi^{\ast} \,]\,
      [\, \partial_\mu \tpsi \,]
   \bigr \} ,
   \label{e:TW-4a} \\
   H(Q)
   &=
   \!\tint \dd{x} \!
   \bigl \{ 
       |\partial_x \tpsi |^2
       -
       \frac{g \, |\tpsi|^{2\kappa+2}}{\kappa + 1}
       +
       V_1(x) |\tpsi|^2 
   \bigr \} ,
   \label{e:TW-4b} \\
   R_{\mu}(Q)
   &=
   \rmi
   \tint \dd{x} 
   V_2(x) \,
   \bigl \{ \,
      [\, \partial_\mu \tpsi^{\ast} \,] \, \tpsi
      -
      \tpsi^{\ast} \, [\, \partial_\mu \tpsi \,] \,
   \bigr \} \>.   
   \label{e:TW-4c}
\end{align}
\end{subequations}
Here we have interchanged $\mu \leftrightarrow \nu$ in the definition of $I_{\mu\nu}(Q)$ from their Eq.~(6) \cite{PhysRevE.82.016606}.  So we see that $R_{\mu}(Q) \equiv - w_{\mu}(Q)$ and we find that
\begin{widetext}
\begin{align}\label{e:TW-5}
   f_{\mu\nu}(Q)
   &=
   \partial_\mu \pi_\nu(Q) - \partial_\nu \pi_\mu(Q)
   \\
   &=
   \frac{\rmi}{2} \tint \dd{x}
   \{ \, 
      [\, \partial_\mu \tpsi^{\ast} \, ]\,[\, \partial_\nu \tpsi \, ]
      +
      \tpsi^{\ast}\,[\, \partial_\mu \partial_\nu \tpsi \, ]
      - 
      [\, \partial_\mu \partial_\nu \tpsi^{\ast} \,] \, \tpsi
      -
      [\, \partial_\nu \tpsi^{\ast} \,] \, [\, \partial_\mu \tpsi \,]
      -
      [\, \partial_\nu \tpsi^{\ast} \, ]\,[\, \partial_\mu \tpsi \, ]
      -
      \tpsi^{\ast}\,[\, \partial_\nu \partial_\mu \tpsi \, ]
      \notag \\
      & \qquad\qquad
      +
      [\, \partial_\nu \partial_\mu \tpsi^{\ast} \,] \, \tpsi
      +
      [\, \partial_\mu \tpsi^{\ast} \,] \, [\, \partial_\nu \tpsi \,]      
   \,\} \>,
   \notag \\
   &=
   \rmi \tint \dd{x}
   \{ \, 
      [\, \partial_\mu \tpsi^{\ast} \, ]\,[\, \partial_\nu \tpsi \, ]
      -
      [\, \partial_\nu \tpsi^{\ast} \,] \, [\, \partial_\mu \tpsi \,] 
   \,\}
   =
   I_{\mu\nu}(Q) \>.
   \notag
\end{align}
\end{widetext}
In the notation used in the variational method, Eq.~\ef{e:TW-3} becomes
\begin{equation}\label{e:TW-6}
   f_{\mu\nu}(Q) \, \dot{Q}^\nu
   =
   u_{\mu}(Q) - w_{\mu}(Q)
   =
   v_{\mu}(Q) \>.
\end{equation}
So the generalized traveling wave approximation is identical to the variational method.  The authors of Ref.~\cite{PhysRevE.82.016606} proved this in another way in Sec.~III of their paper for a simpler dissipative system.

%
%
\bibliography{johns.bib}
%
%
%
\end{document}